\documentclass[showpacs,aps]{revtex4}
\usepackage{epsfig}

\begin{document}
\newcommand{\highlight}{\bf}
\newcommand{\change}{\bf}
\newcommand{\greg}{\bf}

\title{Kondo effect and STM spectra through
 ferromagnetic nanoclusters}
\author{Gregory A. Fiete$^1$, Gergely Zarand$^1$, Bertrand
  I. Halperin$^1$ and Yuval Oreg$^{2}$}
\address{
$^1$Department of Physics, Harvard University, Cambridge,
Massachusetts 01238\\$^2$Weizmann Institute of Science, P.O.Box 26
Rehovot 76100, Israel}
\date{\today}

\begin{abstract}
{ Motivated by recent scanning tunneling microscope (STM) experiments
on cobalt clusters adsorbed on single wall metallic nanotubes
[Odom {\em et al.}, Science {\bf 290}, 1549 (2000)],
we study theoretically the size dependence of 
STM spectra and  spin-flip  scattering of 
electrons from  finite size 
ferromagnetic clusters adsorbed on  metallic surfaces. 
We study  two models of nanometer size ferromagnets: (i) An itinerant
model with delocalized s, p and d electrons and (ii) a local moment model
with both localized  d-level spins and delocalized cluster electrons. 
The effective exchange coupling between the spin of the cluster and the 
conduction electrons of the metallic substrate depends 
on the specific details of the single particle density of 
states on the cluster.  
The  calculated Kondo coupling is inversely proportional to the 
total spin of the ferromagnetic cluster in both models 
and thus the Kondo temperature  is rapidly  suppressed as the { size} of 
the cluster increases.  
Mesoscopic fluctuations in the charging energies and magnetization of 
nanoclusters can lead to large fluctuations
in the Kondo temperatures and  a very asymmetric voltage dependence of
the STM spectra. We compare our results to the experiments.} 
\end{abstract}
\maketitle

\section{Introduction}
Kondo physics\cite{hewson} has seen a revival in recent years due 
to an exciting series of new experiments on mesoscopic\cite{david,leo} and 
nanoscale\cite{li,hari,cobden,teri} systems that has enabled a more
thorough and controlled study of the basic problem of
a local moment interacting with a sea of conduction electrons.  It is
now possible, for example in quantum dots, to tune the Kondo
temperature, $T_K$, with externally applied gate
voltages\cite{david,leo,cobden} or, with the scanning tunneling
microscope (STM), to study the Kondo
effect at single magnetic impurities on metallic
surfaces.\cite{li,hari}  The superb spatial resolution of the STM
permits  unprecedented direct local spectroscopic detail of Kondo impurities. 

Recent low temperature($\sim 4$ K) STM experiments\cite{teri} have
probed the spectroscopy of isolated nanometer and
sub-nanometer ferromagnetic (Co)
clusters on metallic single walled carbon nanotubes.\cite{noluttinger}  
The tunneling spectrum of
these small ferromagnetic clusters exhibited several interesting
features--most notably Kondo effect in sub-nanometer diameter clusters
and well resolved discrete level spacing in nanometer size clusters.  
Both sub-nanometer and nanometer size clusters
exhibited a Coulomb charging gap in the spectrum near zero bias.
The Kondo effect in the sub-nanometer size clusters appeared in the
spectrum as a Fano-like sharp {\it peak} \cite{peak_comment,ujsaghy}with a 
half-width $\sim
15$ meV ($T_K \sim$ 80 K) at 4 K.  This peak was not present at 100 K,
presumably due to thermal destruction of the Kondo resonance. 

The goal of this paper is to present a theoretical framework to study
the spin-flip scattering of conduction electrons from finite size ferromagnetic
clusters adsorbed on  metallic surfaces. { In particular, we are 
 interested in understanding some of the {\em size-dependent features} 
observed in the tunneling spectrum.  We focus on how the Kondo Effect, 
the level spacing and the level widths depend on the size of the cluster.
} 

Cluster ferromagnetism may be described by two alternate models:
(i)  In the experiments of Odom {\it et al.}\cite{teri} magnetism of 
the cluster is most likely {\em itinerant} in nature as bulk Co is
an itinerant band ferromagnet. The basic
physics of itinerant clusters is captured in the model of 
Refs.~\onlinecite{canali,kleff} (see Sec.~\ref{sec:itin}).
(ii) In some other cases, however, 
including the case of many semiconducting ferromagnets
\cite{semiferro} and rare earth materials, ferromagnetism is better
described in terms of {\em local moments}  that couple ferromagnetically or 
antiferromagnetically to the cluster conduction electrons (holes). 
This motivated us to introduce another  exactly solvable model, 
where d-electrons produce highly  localized magnetic moments.
While the excitation spectrum of the  two models turns out to be very 
similar, the way they couple to the  metallic substrate is rather different. 
The most appropriate model to use in a given physical situation depends 
on the material that composes the
nanoparticle and whether it is more appropriate to think of the magnetism of 
the particle as due to localized moments or  itinerant electrons that, although
free to move about, are nevertheless polarized strongly enough to give a net 
spin to the cluster.

{
We make several assumptions during our computations: 
We assume the clusters to have between, say, eight and forty atoms 
and that they
are large enough that bulk properties, such as the magnetization per atom,
are not changed substantially due to their finite size.
Furthermore, we assume that the single particle states on the island
are of extended character, and therefore the single particle level 
spacing decreases as $\sim 1/N_A$ with an increasing number of 
cluster atoms, $N_A$. We also assume that the ferromagnetic 
spin splitting $\Delta_s$ 
of the  extended states on the island is much larger than the single 
particle level  spacing and is approximately independent of the size 
of the cluster.  Thus the total spin of the cluster is rather large and is 
roughly proportional to $\sim N_A\gg 1$.}

Depending on the strength of the electron tunneling  between the 
metal and the cluster we may distinguish two regimes: weak and strong. 
For weak
tunneling we derive an effective Kondo Hamiltonian in second order 
perturbation theory in the tunneling.
In this regime we find that at very low energy scales 
the cluster behaves  most similarly to a quantum dot with a single
unpaired electron.  However, ferromagnetic interactions  
induce a large spin on the cluster and at the same time 
{\em reduce}  the value of  the effective exchange coupling between the 
spin of the cluster and the electrons in the metal.

The sign of the exchange coupling depends on the 
specific details  of the band structure on the nanoparticle. For a 
Co cluster,  in particular,  the itinerant model gives a tendency for 
{\em antiferromagnetic} coupling, and possibly produces a Kondo effect. 
However, the Kondo temperature $T_K$, decreases very fast with increasing 
cluster sizes,  and therefore the Kondo effect can be observed only 
for very small clusters.

The effective couplings we obtain,  $J^{\rm eff}$, turn out to be  in general 
$M \times M$ matrices describing spin-flip scattering among the $M$ 
conduction band orbital modes that couple to the cluster
($M$=4 for metallic nanotubes\cite{dresselhaus}).  
Therefore we expect that a series of Kondo effects  
takes place at Kondo temperatures  corresponding to 
the antiferromagnetic  eigenvalues of $J^{\rm eff}$.
The total spin of the impurity, $S_T$ gets successively screened at 
each Kondo temperature as the  temperature is lowered: 
$S_T \to S_T-1/2 \to S_T-1 \ldots\mbox{ }$.
For a nanotube this compensation can never be complete\cite{nozieres}
for any spin greater than 2  since there are only four channels available for
screening.\cite{dresselhaus} The degeneracy of the remaining unscreened spin 
will ultimately be lifted by magnetic anisotropy induced by 
 spin-orbit coupling.\cite{ujsaghy_2}

To make  closer contact with the experiments of
Odom {\it et al.}\cite{teri} we compute the STM tunneling spectrum
using  the itinerant  model. The simultaneous observation of the Coulomb 
gap and the discrete levels, whose widths almost equal the level spacing,  
suggest that the clusters investigated are in the intermediate 
tunneling regime.  Nevertheless, carrying out 
perturbative calculations in the weak tunneling limit, we are able to 
reproduce  
the essential features of the STM spectrum {\em above} the 
the Coulomb charging energy. We find that small fluctuations 
in the charging energies can give rise to  very asymmetrical 
STM spectra.

{
From the  STM spectra of the clusters at higher voltages,
we can crudely estimate the Kondo temperature, $T_K$, 
 of a small Co cluster of the size that exhibited a 
Kondo effect in Ref.~\onlinecite{teri}.  
However, our estimates tend to give a $T_K$ that is too small. 
We find that typical mesoscopic fluctuations  in the  charging energies  
of a  ferromagnetic cluster  can considerably increase  the 
Kondo temperature, however, they do not  account for the
difference between our  theoretical  estimates and  
the experimentally  observed $T_K$.~\cite{teri}
Nevertheless, the fact that   some small Co clusters exhibit 
Kondo effect while others do not, indicates
that these mesoscopic fluctuations may indeed play an important 
 role in  determining $T_K$: Since $T_K$ depends exponentially 
on $J^{\rm eff}$, relatively small changes in the latter 
can induce large variations of $T_K$. 
}

This paper is organized as follows.  In Sec.~\ref{sec:itin} we 
introduce the itinerant model  of nanoscale ferromagnets. 
In Sec.~\ref{perturbativelycoupled} we study the spin-flip scattering 
of conduction electrons from an itinerant 
 cluster using second order perturbation  theory in the cluster-substrate
tunneling, and we derive a general 
expression for the matrix of Kondo couplings,  $J^{\rm eff}$.
In Sec.~\ref{connection} we make contact to the experiments of Odom
{\it et al.}\cite{teri} by calculating in a simple approximation the
STM tunneling spectrum of a ferromagnetic Co cluster on a metallic
substrate in the limit of weak tunneling. 
In Sec.~\ref{localmoment} we introduce a new local moment 
model for ferromagnetism and show how to compute
the effective interaction  $J^{\rm eff}$ in this case. 
{In Sec.~\ref{discussion}
we discuss in detail the connection between our work and recent experiments.} 
Finally, in Sec.~\ref{conclusions} we present our conclusions.

\section{Itinerant Ferromagnetic Cluster Model}
\label{sec:itin}
 
\begin{figure}
\begin{center}
\epsfig{figure=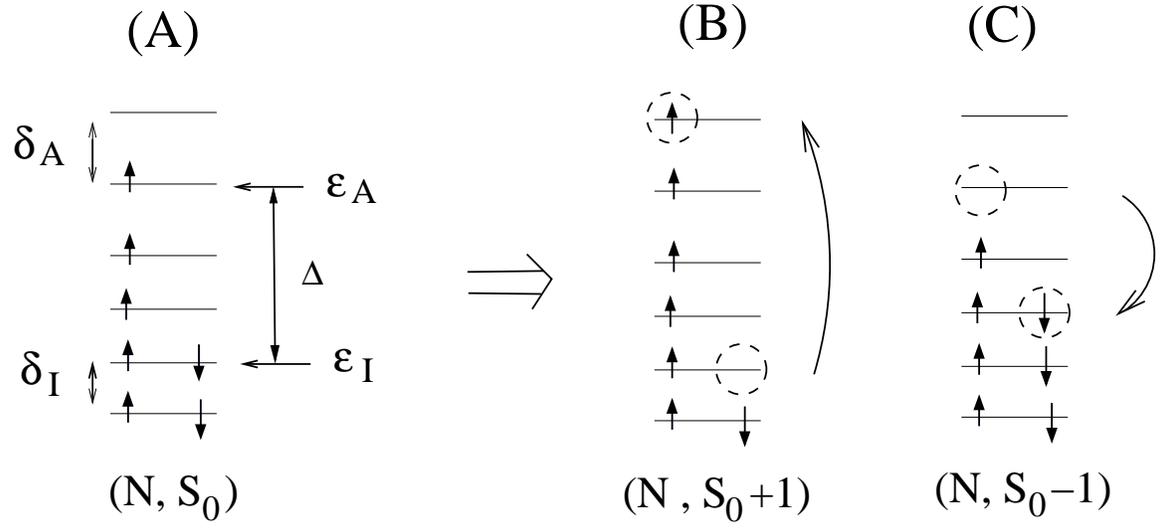,width=6in}
\end{center}
\caption{Spin excitations of a  ferromagnetic nanoparticle.
For the precise definition of $\epsilon_{A/I}$,  $\delta_{A/I}$, and
$\Delta_s$ see text. 
Fig.~A: The fully porarized  ground state. 
Figs.~B and C: Lowest lying particle-hole excitations having energy 
$\sim\delta_{A/I}$. 
}
\label{states}
\end{figure}

An itinerant mean field model of a magnetic nanoparticle 
was first proposed by Canali and
MacDonald\cite{canali} and later extended by Kleff {\it et al.}\cite{kleff} 
to explain the dense spectrum in tunneling
measurements of nanometer size Co particles.\cite{ralph}   The basic
Hamiltonian is 
\begin{equation}
\hat H_{\rm cluster}=\sum_{j,\sigma}\epsilon_j c^\dag_{j \sigma}c_{j \sigma}
  -{1\over2} {U\over 
  N_A}{\vec S} \cdot {\vec S}  + {E_C\over 2}({\hat N}-n_g)^2
\label{model_I}
\end{equation}
where  $c^\dag_{j \sigma}$ ($c_{j \sigma}$) creates (destroys)
an electron with spin $\sigma$ at the $j^{th}$ energy level
of the cluster with kinetic energy  $\epsilon_j$ 
(See Fig.~\ref{states}). The first term in Eq.~(\ref{model_I}) 
represents  the kinetic energy of the 
 hybridized $s, p \mbox{ and }d$-band electrons.
The parameter $U>0$ is the effective
ferromagnetic exchange interaction on the cluster, $N_A$ is the
number of atoms that constitute the cluster, and  
\begin{equation}
\vec S=\sum_j{1\over 2}\sum_{\alpha, \alpha'} c^\dag_{j \alpha} 
\vec \sigma_{\alpha \alpha'}c_{j \alpha'}
\label{spin_sum}
\end{equation}
is the total spin of the cluster. 
 The second term in Eq.~(\ref{model_I}) corresponds to 
ferromagnetic exchange on the cluster: It gives rise to a spontaneous 
polarization of the cluster  by making spin alignment of different 
levels energetically favorable.  
 Finally the last term describes the
charging of the cluster with $E_C$ the Coulomb charging
energy, ${\hat N} = \sum_{j, \sigma}\hat n_{j \sigma}$ the total number 
of electrons on the cluster,  and $n_g$  a dimensionless gate 
voltage that determines the number of electrons in the ground 
state.\cite{quantumdot}  

{ Throughout this paper we are 
neglecting  fluctuations in the level spacing.  
In  a {\em real} nanoparticle, the level spacing between the $j^{th}$
and  $j+1^{th}$ energy level can be written as
$\delta_j=\langle \delta_j\rangle + \eta_j$, 
where $\langle \delta_j\rangle$ is the level spacing corresponding to 
the bulk  energy-dependent density of states,
$\langle \delta_j\rangle\sim 1/\varrho(\epsilon_j)$,  
and $\eta_j$ ($\langle \eta_j \rangle$=0) is 
a fluctuating part due to shape irregularities in the nanoparticle. In this
paper we take $\eta_j \equiv 0$, corresponding to an infinitely strong 
level repulsion.}

The advantage of this model is that  $S^z$, $S^2$ and $n_j$ are conserved
quantities and therefore the Hamiltonian in Eq.~(\ref{model_I}) can be 
diagonalized 
exactly. In the more general situation a magnetic anisotropy term 
should be added to the Hamiltonian.
 Nevertheless, for the
experimentally investigated  
clusters  this anisotropy is estimated to be much less then the width 
of the Kondo resonance
observed, and therefore it can be neglected.\cite{anisotropy_comment}

The ground state  can be explicitly constructed:\cite{kleff}
\begin{equation}
|S_0,S^z = S_0\rangle_G^{N}=\prod_{j=1}^{n_{\uparrow}}c_{j \uparrow}^\dagger
 \prod_{j=1}^{n_{\downarrow}}c_{j \downarrow}^\dagger|vac\rangle
\label{ground_state}
\end{equation}
with $S_0$ and $N$ the spin and particle number  in the ground state. 
The remaining  states within the ground state multiplet 
may be explicitly constructed using the lowering operator:
\begin{equation}
|S_0,S^z\rangle_G^{N}={\sqrt{(S_0+S^z)!\over (2S_0)! (S_0-S^z)!}}
 (S^-)^{(S_0-S^z)}|S_0,S_0\rangle_G^{N}.
\label{spin_lower}
\end{equation}

In itinerant ferromagnets there is an approximate rigid band splitting 
between spin up and spin down
electron density of states.\cite{papa} { Throughout this paper, we 
will assume that our clusters are large enough that the band splitting energy,
$\Delta_s$, is well approximated by the bulk value of the specific material
we are considering.} In our model $\Delta_s$ can be defined as the energy 
difference between the 
highest occupied spin up level, $\epsilon_A$, and the highest 
occupied spin down level, $\epsilon_I$:
$$
\Delta_s \equiv \epsilon_A-\epsilon_I\;,
$$
and { from band structure calculations it is known to be} typically of 
the order of a few electron volts\cite{papa} 
(see Fig.~\ref{states}).
It can be determined by
demanding that the ground state of the
ferromagnetic cluster be stable to fluctuations of energy level
occupations with constant particle number, and is related to the 
interaction parameter $U$ as\cite{canali}
\begin{equation}
U={N_A \over S_0}\Delta_s+d_0\;,
\label{U}
\end{equation}
where $d_0$ is a small quantity that scales as $1/N_A$, 
{ while the first term is roughly independent of the 
cluster size, since $S_0\sim N_A$}. 
Using Eq.~(\ref{U}), one can estimate  the energy cost   
 of a particle-hole excitation (see  Fig.~\ref{states}B,C). 
This turns out to be of the order  of the level 
spacing ($\propto 1/N_A$) (and  {\em not} the exchange  splitting $\Delta_s$):
\begin{equation}
\delta E(S_0\pm1,N)\equiv {\rm min}\{ E_{\rm excited}(S_0\pm1,N)-E_G \}
\sim \delta_A, \delta_I\;.
\label{eq:ph}
\end{equation}
where $E_G$ denotes the ground state energy, and $\delta_A$
($\delta_I$) is the level spacing near   
$\epsilon_A$ ($\epsilon_I$). If $U$ is large enough to fully polarize
the cluster, the energy scales outlined above may
change.\cite{footnote_splitting} 

The minimum cost of adding a particle or a hole to the cluster
can be defined as,
\begin{equation}
\delta E_{\pm,\sigma} \equiv {\rm min}
\bigl\{ E(N\pm1,S_0 +\sigma/2) - E_G  \bigr\},
\end{equation}
where the minimum is over all possible excited states. To determine  
$ \delta E_{\pm,\sigma}$ we consider  processes  like 
those in Fig.~\ref{addition}.   {  These energies can be estimated as
\begin{equation}
\delta E_{+,\uparrow} \approx \bar\epsilon + E_C^+ + \Biggl(\delta_A - {1\over2}{S_0\over N_A} d_0
- {3\over8}{\Delta_s \over S_0} \Biggr)\;,
\label{mu}
\end{equation}
\begin{equation}
\delta E_{+,\downarrow} \approx \bar\epsilon + E_C^+ + \Biggl(\delta_I + {1\over2}{S_0\over N_A} d_0
+ {1\over8}{\Delta_s \over S_0} \Biggr)\;,
\label{mu_2}
\end{equation}
\begin{equation}
\delta E_{-,\uparrow} \approx \bar\epsilon + E_C^- 
- \Biggl( {1\over2}{S_0\over N_A} d_0
+ {3\over8}{\Delta_s \over S_0} \Biggr)
\;,
\label{mu_3}
\end{equation}
\begin{equation}
\delta E_{-,\downarrow} \approx \bar\epsilon + E_C^- + \Biggl( {1\over2}{S_0\over N_A} d_0
+ {1\over8}{\Delta_s \over S_0} \Biggr)\;,
\label{mu_4}
\end{equation}
}
where the corrections scale as $O(1/N_A^2)$, and
$E_C^\pm$ denote the charging energies in the limit of 
vanishing ferromagnetic coupling and an infinitely dense single 
particle spectrum: 
\begin{equation}
E_C^{\pm} \equiv E_C({1\over 2 }\pm (N-n_g))\;.
\end{equation}
The parameter  $\bar\epsilon \equiv (\epsilon_A +\epsilon_I)/2$
in Eq.~(\ref{mu}) can be absorbed into the definition of $n_g$, 
and we will set  it to zero in what follows.

{ The last terms in Eqs.~(\ref{mu}-\ref{mu_4}), 
proportional to $\Delta_s/S_0$, are specific to  the mean field 
model discussed here and have little
physical meaning. Their effect can be fully taken into account
by renormalizing the `chemical potential' $\bar \epsilon$ and 
the mesoscopic parameter $d_0$, and they therefore do
not modify our results.}

Using a stability analysis similar to the previous one  it is a trivial 
matter to show that typically
\begin{equation}
\delta E_{\pm,\sigma}\sim E_C/2\;.
\end{equation}

For later purposes it is also useful to have excitation energies of
particles and holes added to {\em any} level defined, $\delta
E^j_{\pm,\sigma}(N\pm1,S_0 + \sigma /2)$. For example, 
increasing the total spin by adding an electron to the 
$j^{th}$ unoccupied level  rather than  the one immediately 
above $\epsilon_A$ gives, 
\begin{equation}
\delta E_{+,\uparrow}^j \equiv 
\epsilon_j-(\epsilon_A+\delta_A)
+ \delta E_{+,\uparrow}\;.
\label{energy_1}
\end{equation}
Removing an electron from level $j$ below $\epsilon_I$ 
so that the total spin increases gives
\begin{equation}
\delta E_{-,\uparrow}^j \equiv \epsilon_I -\epsilon_j
+\delta E_{-,\uparrow}\;.
\label{energy_2} 
\end{equation}

It is important to point out that in our model all effects of mesoscopic
fluctuations have been put into the quantities $E_C$, $d_0$ and $n_g$.  
All other
quantities of the itinerant model are specified by the bulk band structure
and the number of atoms, $N_A$, in a given nanocluster. 

\section{Weak Tunneling between a Ferromagnetic Cluster and a Metal}
\label{perturbativelycoupled}

\subsection{General Expression}
A finite size ferromagnetic cluster can  scatter
electrons between different orbital channels in the substrate
and flip their spin while doing so.
If the spin-orbit coupling is sufficiently weak then $SU(2)$ symmetry in 
the spin sector implies that the low-energy effective 
 interaction between the cluster spin and the electrons takes the 
form (apart from a potential scattering term):
\begin{equation}
\hat H_{\rm Kondo}^{\rm eff} ={1\over 2} \sum_{k,k'}J^{\mu\nu}\vec S \cdot
a^\dagger_{\mu k \alpha}\vec
\sigma_{\alpha \alpha'} a_{\nu k' \alpha'}
\label{eq:Kondo}
\end{equation}
where $J^{\mu\nu}$ is a Hermitian matrix of spin-flip 
exchange couplings among
the various orbital channels, indexed by $\mu$ and $\nu$. 
(In a one-dimensional system, such as a carbon nanotube, $\mu$ labels 
left-going and right-going modes of different symmetries. 
In an isotropic two or three dimensional host $\mu$ labels
angular momenta about the cluster.)
The total spin of the cluster is $\hat S$ and the $a^\dagger_{\mu k \alpha}$
($a_{\mu k \alpha}$) are creation (annihilation) operators for  
conduction electrons of the metal in orbital channel $\mu$,
wavenumber $k$ and spin $\alpha$.  A finite size cluster in an isotropic host
will generally scatter electrons among all orbital channels, 
but in practice
$J^{\mu\nu}$ can be truncated to include a finite number of channels,
the largest angular momentum being $l\sim {L\over \lambda_F}$ where $L$ is the
size of the cluster and $\lambda_F$ is the Fermi wavelength of the
conduction electrons.

In the limit of weak tunneling, $J^{\mu\nu}$ can be calculated in
second order perturbation theory in the tunneling.  
The Hamiltonian we consider is $\hat H=\hat H_0 + \hat V$, 
where $\hat H_0 =\hat H_{\rm metal} + \hat H_{\rm cluster}$ and 
where $\hat H_{\rm metal}=\sum_{\mu, k,\sigma} \epsilon_{\mu,k}
a^\dagger_{\mu  k \sigma}a_{\mu  k \sigma}$ is the Hamiltonian of 
the free conduction electrons of the metal and
\begin{equation}
\hat V =\sum_{\mu,\sigma,j,k}(V_\mu^{j,k}c_{j \sigma}^\dagger
  a_{\mu k \sigma} + h.c.)\;,
\label{V_use}
\end{equation}
where
\begin{equation}
V_\mu^{j, k} \equiv \sum_{n=1}^{N_p} V_n \varphi_j^*(\vec R_n)
\psi_{\mu,k}(\vec R_n) \;,
\label{V_n_sum}
\end{equation}
and $N_p$ is the number of points of electrical contact
between the cluster and the metal.
The $\varphi_j(\vec R_n)$ are the wave functions of the $j^{th}$
levels on the cluster at the points of contact $\vec R_n$ and the
$\psi_{\mu,k}(\vec R_n)$ are the wave functions of the metal at $\vec R_n$. 
The tunneling amplitude $V_n$ can be estimated with a knowledge of 
the wave functions near
the cluster-metal contact points and the work functions of the cluster
and the metal.  

We can determine $J^{\mu\nu}$ by equating
\begin{equation}
\langle f|\hat H_{\rm Kondo}^{\rm eff}|i\rangle =\sum_n {\langle f|\hat
  V|n\rangle \langle n|\hat V |i\rangle \over
  E_i -E_n}\;,
\label{pert}
\end{equation}
where 
$|i\rangle =|S_0,S^z\rangle _G^{N}|\nu,k_i,\sigma\rangle$
and 
$|f\rangle=|S_0,S^{z'}\rangle_G^{N}|\mu,k_f,\sigma'\rangle$
denote the initial and final states with energies 
$E_i$ and $E_f$, respectively, and $|n\rangle$ stands for all possible 
 intermediate states with energy $E_n$, $H_0|n\rangle =E_n|n\rangle$.
Here  $|\nu,k,\sigma\rangle$ denotes the  Fermi sea with an 
additional electron in the  $\nu$th channel with momentum 
$k$ and spin $\sigma$.
  
\begin{figure}
\centerline{\epsfig{figure=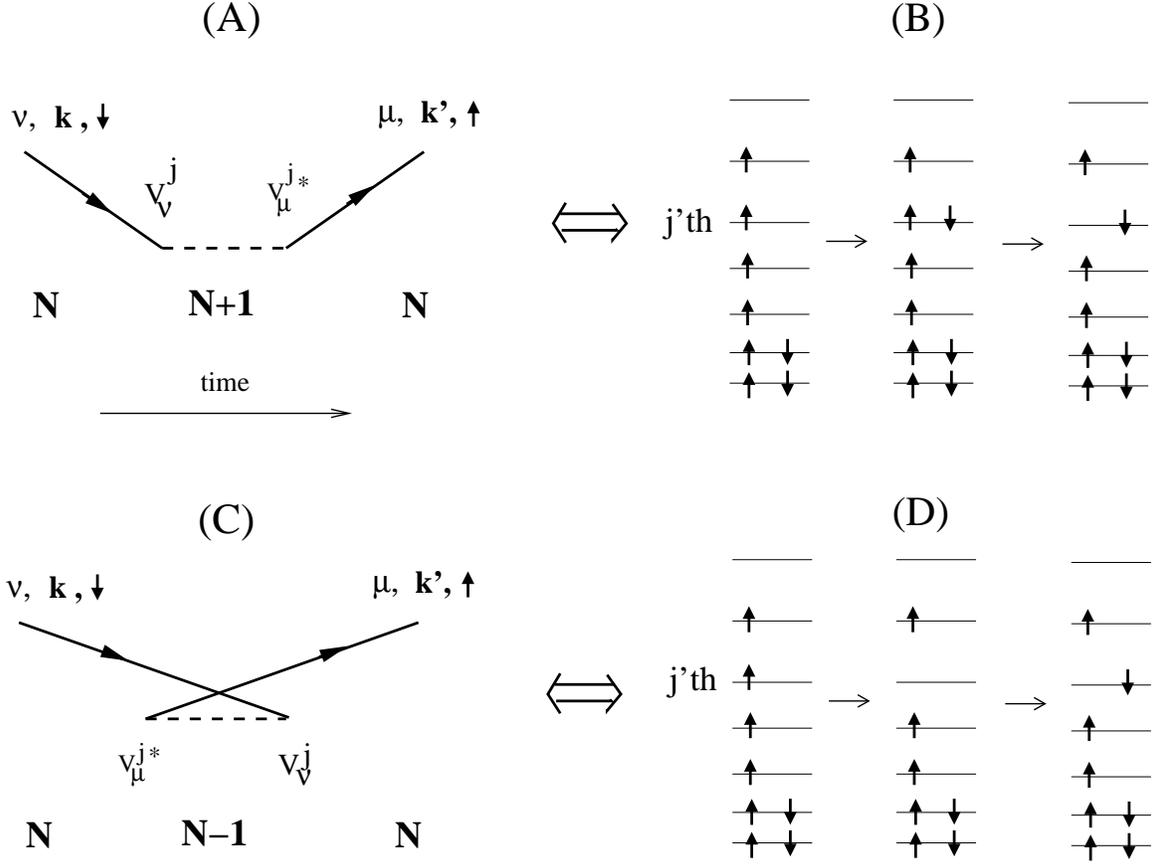, width=6 in}}
\caption{Kondo scattering from a ferromagnetic cluster.
Figs.~A and C: Time ordered diagrams 
in  second order perturbation generating 
the single particle contribution Eq.~(\ref{J_s}) to the
effective  exchange  $J^{\mu \nu}$.
Continuous lines represent the incoming and outgoing electrons
while dashed lines denote the intermediate excited state of the 
nanoparticle.
In A an  electron in state $|\nu,k,\downarrow>$
  hops on the cluster onto level $j$ and  another
  electron leaves the cluster from level $j$ with outgoing state
  $|\mu,k',\uparrow>$. 
In C first an electron hops out from level $j$  into state
  $|\mu,k',\uparrow>$ and then  the incoming electron  hops onto 
  level $j$.  Level occupation changes are shown in D. 
}
\label{kondo}
\end{figure}

The easiest way to determine $J^{\mu\nu}$ is to focus on purely
 spin-flip process to which potential scattering does not contribute, and 
choose $S^z=S_0, \sigma=\downarrow$ and $S^{z'}=S_0-1, \sigma'=\uparrow$, 
giving $\langle f|\hat H_{\rm Kondo}^{\rm eff}|i\rangle={1\over 2}\sqrt{2S_0} 
J^{\mu  \nu}$. The RHS of Eq.~(\ref{pert}) is evaluated by summing over both
states with an extra particle and states with an extra hole in the 
intermediate state  (See  Fig.~\ref{kondo}). After a rather 
straightforward computation we find that $J^{\mu\nu}$ is given by the sum of 
three contributions:
\begin{equation}
J^{\mu\nu} = J^{\mu\nu}_d +J^{\mu\nu}_s + J^{\mu\nu}_e\;.
\label{eq:J_sum}
\end{equation} 
The couplings $J^{\mu\nu}_d$ and  $J^{\mu\nu}_e$  describe the 
contributions from tunneling processes involving  doubly occupied and
empty single particle levels: 
\begin{eqnarray}
J_d^{\mu \nu} &=& - {1 \over S_0+1/2} 
\sum_{j \le I }  {V_\mu^{j,k_f}}^* V_\nu^{j,
  k_i}
\Biggl({1 \over \delta E_{-,\uparrow} +\epsilon_{I} -\epsilon_j}  - 
{1 \over \delta E_{-,\downarrow}+\Delta_s +\epsilon_{I} -\epsilon_j} 
  \Biggr)\;\,\label{J_d}\\
J_e^{\mu \nu} &=& - {1 \over S_0+1/2} 
\sum_{j> A }  {V_\mu^{j,k_f}}^* V_\nu^{j,
  k_i}
\Biggl({1 \over \delta E_{+,\uparrow} -\epsilon_{A+1} +\epsilon_j}  - 
{1 \over \delta E_{+,\downarrow} + \Delta_s -\epsilon_{A+1} +\epsilon_j} 
  \Biggr)
\label{J_e}\;,
\end{eqnarray}
and the incoming and outgoing electrons were put at the Fermi energy.
Since $\Delta_s \gg \delta E_{\pm,\sigma}$
both contributions are {\em ferromagnetic}, {\em i.e.}, all eigenvalues 
of $J^{\mu\nu}_d + J^{\mu\nu}_e$ are negative. To see this 
let us  first neglect the $k$-dependence of the $V_\mu^{j,k}$'s. 
Then the matrices $J^{\mu\nu}_a$ ($a=e,d$) can be expressed as a sum:   
$J_a^{\mu\nu}= \sum_{j} P^{\mu\nu}_{j}$. Each of these terms 
is obviously negative semidefinite, 
{\em i.e.} for any vector $a^\mu$ the product  $a^*_\mu P_{j}^{\mu\nu}a_\nu$
is negative or zero. As a consequence,  $J_{e}^{\mu\nu}$ and
$J_{d}^{\mu\nu}$ are also negative  semidefinite, {\em i.e.}, 
in a diagonal  basis all their eigenvalues  are negative or zero. This 
simple proof can readily be  extended for k-dependent  $V_\mu^{j,k}$'s.

{ The physical reason that the doubly occupied and empty single 
particle states give rise to a ferromagnetic contribution to 
$J^{\mu \nu}$ is that
in the intermediate state in second order perturbation theory  
an electron (hole) that hops onto the island
with spin parallel to that of the cluster has smaller energy than 
than an electron (hole) with opposite spin orientation,
due to the ferromagnetic exchange {\em on} the cluster. 
%(antiparallel) 
Therefore an electron (hole) on the substrate with spin parallel 
to that of the cluster can lower its kinetic energy more efficiently
through hopping to the empty (doubly occupied) 
states. This is reflected by the smaller 
energy denominator in Eqs.~(\ref{J_d}) and (\ref{J_e}).}

Singly occupied levels, on the other hand, give an 
{\em antiferromagnetic} contribution 
to $J^{\mu \nu}$: 
\begin{equation}
J_s^{\mu \nu} = {1 \over S_0} \sum_{j=I+1}^A {V_\mu^{j,k_f}}^* V_\nu^{j,
  k_i}\Biggl({1 \over 
\delta E_{+,\downarrow} -\epsilon_{I+1} +\epsilon_j}  + 
{1 \over \delta E_{-,\downarrow}+\epsilon_A -\epsilon_j}
\Biggr), 
\label{J_s}
\end{equation}
with $J_s^{\mu \nu}$ having only non-negative eigenvalues.\cite{Glazman} 
This is a result of the Pauli principle,  because only 
electrons (holes) in states with 
spin antiparallel 
%(parallel) 
to that of the cluster may hop 
onto singly occupied levels in second order perturbation 
theory, and thereby reduce their kinetic energy.

Eqs.~(\ref{J_d}),   (\ref{J_e}), and (\ref{J_s}) constitute
 one of the central results of the paper. They   describe how 
$J^{\mu \nu}$ depends on the density of states of the
singly occupied levels on the cluster, the excitation energies to the
$N \pm 1$ manifold of states and the tunneling amplitudes to the
various levels of the cluster.

In general, there is a competition among the three terms of 
Eq.~(\ref{eq:J_sum}), and thus the sign of  eigenvalues of 
$J^{\mu\nu}$ depends on the specific structure of the 
single particle density of states on the cluster, and
other  mesoscopic parameters such as $\delta E_{\pm,\sigma}$. 
In the {\em absence} of ferromagnetic interactions on 
the cluster, $U=0$, however, 
the ferromagnetic contributions $J^{\mu\nu}_e$ and $J^{\mu\nu}_d$
identically vanish, and the effective interaction is always 
antiferromagnetic, provided there is an odd number of electrons 
on the  cluster.

{ 
The  itinerant model has been used independently
in Ref.~\onlinecite{Glazmanpaper}  to describe 
spin $S=1$ metallic nanoclusters 
with similar results. In that case the  
ferromagnetic interaction is  weak,  
$U < \delta$,   the island is far from a ferromagnetic unstability, and 
the   $S=1$ ground state of the island results rather from  
two  single particle states of the island having 
energies accidentally  closer than $U <  \delta$. 
Then the ferromagnetic contribution of the doubly occupied and unoccupied
levels is usually negligible, and the effective exchange interaction 
is dominated by the antiferromagnetic contributions of the 
singly occupied levels.\cite{Glazmanpaper}
}

The off-diagonal elements of $J^{\mu\nu}$ are a sum of random numbers since 
the $\varphi^*_j(\vec R_n)\psi_{\mu,k}(\vec R_n)$ of Eq.~(\ref{V_n_sum}) have
random phase for different $j$ and $\mu$'s.
Therefore  the size of the  off-diagonal elements of  
$J^{\mu \nu}$ will be down by a factor of $ \sim \sqrt{1 \over  2 S_0} 
\sim \sqrt{\langle \delta_{A,I}\rangle  \over \Delta_s}$ compared 
to the diagonal ones with $\mu=\nu$, and  the matrix $J^{\mu\nu}$ will 
be dominated by the  diagonal elements if the number of scattering 
channels  $M$  is much  less  than $2 S_0$. 

\subsection{Weak Tunneling at a Single Point of Contact}
\label{onepoint}
It is instructive and also useful to  specialize the results of 
Sec.~\ref{perturbativelycoupled}A
to  the important case of a ferromagnetic cluster that makes
contact with a metallic substrate in only a single point.  Then we can choose
a basis where $J^{\mu \nu}$ consists of a single non-zero element as
all but one conduction channel will decouple from the cluster { (that is,
a linear combination of the host orbital modes can be found such that only
one orbital mode in the new basis has non-zero amplitude at the impurity 
location)}.  
With the origin taken as the single point of contact, $V_\mu^{j, k}$ 
of Eq.~(\ref{V_n_sum}) reduces to $V_j \equiv V \varphi_j^*(0)\psi_{k}(0)$.
Eq.~(\ref{J_s}), {\em e.g.},  then becomes
\begin{equation}
J_s=   \sum_j {|V_j|^2 \over S_0}\Biggl({1 \over 
\delta E_{+,\downarrow} -\epsilon_{I+1} +\epsilon_j}  + 
{1 \over \delta E_{-,\downarrow}+\epsilon_A -\epsilon_j}
  \Biggr).
\label{J_single}
\end{equation}
In the limit $\Delta_s,\delta E_{\pm,\sigma} \gg \delta$   
Eqs.~(\ref{J_e}), (\ref{J_d}), and (\ref{J_s}) can be  approximately 
evaluated by assuming that the tunneling matrix amplitudes do not 
vary too much from level to level,  and replacing the sums by integrals. 
For $J_s$, {\em e.g.}, we obtain:
\begin{equation}
J_s \approx {\langle|V_j|^2\rangle_j\over S_0} \int_0^{\Delta_s}  d\xi
\varrho (\epsilon_I + \xi)  
\Biggl({1\over \delta E_{+,\downarrow}+ \xi} + {1\over  \delta E_{-,\downarrow}+
\Delta_s -  \xi}\Biggr)\;,
\label{eq:J_est_int}
\end{equation}
with $\varrho (\epsilon)$ the single particle density of states 
on the cluster, and  $\langle|V_j|^2\rangle_j$ the average hybridization 
strength.  
The ferromagnetic couplings $J_d$ and  $J_e$ can be 
expressed by  similar integrals. 
For $S_0 \gg 1/2$ these integrals can be combined in a straightforward 
 calculation where the integrand ${1 \over \delta E_{\pm, \sigma}
+ \xi}$ is separated into the two regions $\xi \gg \delta E_{\pm, \sigma}$ 
and $\xi \ll \delta E_{\pm, \sigma} $ and then approximated in each region.
We obtain the following estimate for the effective exchange coupling valid
when $\delta E_{\pm,\sigma} \gg \delta$: 
\begin{equation}
J^{\rm eff} \sim {\langle|V_j|^2\rangle \over S_0 }
\Biggl[P \int_{-\infty}^\infty {\Delta_s \varrho(\xi) d\xi \over (\xi -\epsilon_I)
(\epsilon_A -\xi)}
-\varrho(\epsilon_I) \mbox{ln} \Biggl({\delta E_{+,\downarrow}  \over \delta E_{-,\uparrow}} \Biggr)
+\varrho(\epsilon_A) \mbox{ln} \Biggl({\delta E_{+,\uparrow}  \over \delta E_{-,\downarrow}} \Biggr)\Biggr]\;.
\label{eq:Jestimate}
\end{equation}
This simple expression is one of the  central results of this work.  
Eq.~(\ref{eq:Jestimate})
describes the dependence of $J^{\rm eff}$ on the {\em particular 
details of the density of states} of an itinerant ferromagnetic nanocluster.  
The first term involves a principal value integral (denoted by $P$) 
over energy and
clearly shows that whenever $\epsilon_I < \xi < \epsilon_A$ the 
contribution to $J^{\rm eff}$ is  antiferromagnetic; otherwise it is
ferromagnetic, {\em i.e.} electron and hole excitations in the singly occupied
states give rise to an antiferromagnetic coupling while electron (hole) 
excitations of the empty (doubly occupied) states give rise to a ferromagnetic
contribution. The last two terms of Eq.~(\ref{eq:Jestimate}) are 
{\em mesoscopic fluctuations} that
depend on the specific charging energies of the nanocluster and the density of
states at the top of the minority and majority bands.  { Other mesoscopic 
fluctuations come from variations in the tunneling matrix elements and 
in the level spacing, which are ignored in our model. } These mesoscopic 
corrections  become more pronounced for smaller 
cluster sizes and lead to strong fluctuations around  $J^{\rm eff}$.  

For Co, the single particle density of states has 
a maximum within the spin-polarized  part of the spectrum,
$\epsilon_I < \xi < \epsilon_A$, resulting in
a $J^{\rm eff}$ that tends to be  {\em antiferromagnetic}.
Numerical evaluation of Eq.~(\ref{eq:Jestimate}) using the actual 
density of states for Cobalt~\cite{papa} shows that the ferromagnetic 
contributions to $J^{\rm eff}$ from $J_d$ and $J_e$ reduce the
dominant antiferromagnetic contribution from $J_s$ by roughly 50-60 percent.
Fluctuations of $\delta E_{\pm,\sigma}$ in small clusters by a factor of 
2 can lead to sample to sample
fluctuations of $\sim$10\% in $J^{\rm eff}$ for Co.  This ultimately leads
to large fluctuations in the Kondo temperature, $T_K$, from cluster to
cluster, since $J^{\rm eff}$ appears in the exponent 
of the expression for $T_K$.

 Eq.~(\ref{eq:Jestimate}) also describes  how $J^{\rm eff}$ 
scales with cluster size.  From the normalization of the cluster 
wave functions we find     $\langle |V_j|\rangle^2\sim 
{1\over N_A}$. { Assuming that the magnetization of the cluster
per atom is determined  by microscopic mechanisms
and thus $U\sim \Delta_s$ does not depend on the cluster size, both 
$S_0 \propto N_A$ and $\varrho \propto N_A$. This results in 
a fast decrease of $J^{\rm eff}$ with increasing cluster size (spin), 
 $J^{\rm eff} \propto {1 \over N_A} \propto {1 \over S_0}$, 
and therefore an exponentially suppressed $T_K$.} Note that  
this suppression in $T_K$ {\em depends only on 
the level structure} of the ferromagnetic nanocluster, not on any 
interference effects which may come from  several points of 
contact.\cite{levin}

{ It is useful at this point to mention what would happen if 
$U\ll \delta_{I,A}$ and we had an odd number of electrons on the cluster.
There is then only one singly occupied level on the cluster, the sums 
in Eqs.~(\ref{J_d}) and (\ref{J_e}) vanish, and the
sum in Eq.~(\ref{J_s}) reduces to just two terms.  As a result, for the
spin 1/2 cluster,
 $J_{1/2}^{\rm eff} \propto {\langle |V_j|\rangle^2 \over E_C} \propto 
{1 \over N_A^{1-\alpha}}$ with $\alpha>0$ since  $E_C \propto 1/N_A^\alpha$.
Thus, for sufficiently large $N_A$, $J^{\rm eff}$ for the ferromagnetic
cluster ($U\gg \delta_{I,A}$) will scale to zero faster with $N_A$ than 
for the non-magnetic cluster.
In Sec.~\ref{sec:estimate_kondo_temp} and in the conclusions,  we will 
comment on what this means for the scaling of the Kondo temperature 
with the number of atoms in such a cluster.}

%GERGELY: I AM NOT SURE WHAT TO DO WITH THIS PART--REMOVE?
%The scaling of $E_C={e^2 \over
%  C_\Sigma}$, the Coulomb charging energy of the cluster, can be estimated
%very crudely as $E_C \sim {1\over L^2}$ for a pancake-shape cluster or
%$E_C \sim {1\over L}$ for a more spherical cluster, where $L$ is the
%size of the cluster.  
%It is instructive to compare  Eq.~(\ref{eq:Jestimate}) with the exchange 
%coupling we would obtain in the {\em absence} of ferromagnetic coupling 
%$U$,  assuming there is one singly occupied level on the cluster:
%\begin{equation} 
%J^{\rm eff}_{\rm single} \sim {V^2 \over E_C/2}.
%\label{anderson}
%\end{equation}  
%We see that the  logarithm in $J^{\rm eff}$ is a new feature of the 
%ferromagnetic cluster which arises from the presence of  multiple 
%levels. Neglecting  the ferromagnetic contribution of the 
%unoccupied and doubly occupied levels we obtain:
%\begin{equation}
%{J^{\rm eff}_{\rm ferro} \over J^{\rm eff}_{\rm single}} 
%\sim {E_C/2\over \Delta_s }
%\mbox{ln} \Biggl({\Delta_s \over E_C/2} \Biggr) <1. 
%\end{equation}
%Thus the ferromagnetic coupling  $U$ leads to a  {\em decrease}  
%in the effective exchange coupling to the magnetic moment of the 
%cluster and to a reduction of  the corresponding Kondo temperature.  
%The ferromagnetic contributions we neglected reduce $J^{\rm eff}$ 
%further. 

\subsection{Weak Tunneling at Several Points of Contact} 

For tunneling at several points of contact we again
turn to Eqs.~(\ref{eq:J_sum}-\ref{J_s}). In this case, $J^{\mu \nu}$ is
necessarily a matrix reflecting the scattering of electrons in more than one
orbital channel. 
As Eq.~(\ref{V_n_sum}) shows, there will be significant random fluctuations 
in the matrix elements $J^{\mu \nu}$ from the $\varphi_j(\vec R_n)$ as
cluster size is changed.  For $N_p$ points of contact there are
a maximum of $N_p$ orbital channels that will be scattered.
However, if $N_p$ is large, the number of orbital channels that are
scattered may be much smaller.  The largest angular momentum (orbital)
channel scattered being $l\sim {L\over \lambda_F}$,  the matrix
$J^{\mu \nu}$ is approximately of size  $\sim l^2 \times l^2 $ 
for a cluster in bulk or $\sim l \times l$ for a cluster 
in contact with a two-dimensional electron gas.  
For a nanotube at most 4 orbital channels can scatter low
energy electrons.\cite{dresselhaus} 

As we discussed earlier 
in this case, in principle at least,  one may observe 
a series of Kondo effects. The cluster displays
an under-screened Kondo effect, where the spin of the impurity is large and
several channels with different coupling constants try to screen it.  The
channel with the largest coupling to the impurity screens half a spin first and
ceases to interact with the impurity. Then the channel with the largest 
coupling from the remaining channels screens half a spin and so on.
In practice, however, the various 
Kondo temperatures are exponentially separated,  and the  
very small Kondo temperatures are likely to be much 
smaller then the spin-anisotropy energy of the cluster. 
Therefore in a realistic  situation 
one can only observe one, or possibly  two Kondo effects.\cite{leo_2}

\section{STM Tunneling Spectra and Kondo Effect}
\label{connection}
\subsection{STM Tunneling Spectra of Ferromagnetic Clusters}
In the simplest model of STM spectra the tip is treated
as a point source of electrons and the resulting signal is
proportional to the local density of states (LDOS) at the 
 tunneling position  $\vec r$,\cite{tersoff}
\begin{equation}
{dI\over dV}(\vec r, V)\propto LDOS(\vec r, \omega = eV) \equiv -{1\over
  \pi}\mbox{Im}[G^R(\vec r,\omega=eV)].
\label{dI_dV} 
\end{equation}
Here $ G^{R}(\vec r,\omega)$ is the retarded Green's function of the
substrate below the tip, and 
$V$ is the voltage difference between the STM tip and the substrate. 
Below we consider only tunneling directly into the
cluster, ignoring any Fano-type interference 
effects.\cite{peak_comment,ujsaghy}  

Let us first focus on the high-energy part of the spectrum
at $\omega > T_K$,  where the Kondo-type strong correlations are negligible.
Using standard manipulations we can express the retarded 
Green's function as 
\begin{eqnarray}
G^R(\vec r,\omega)\approx {1\over N_G} \sum_{G,E,E'}\Biggl(
  \langle G|\hat \varphi(\vec r)|E\rangle
\left({1\over (\omega -(\hat E-E_G)) - M^R(\omega)} \right)_{EE'} 
\langle E'|\hat \varphi^\dagger(\vec r)|G\rangle \nonumber \\
+  \langle G|\hat \varphi^\dagger(\vec r)|E\rangle 
\left({1\over(\omega + (\hat E-E_G)) - M^R(\omega)}\right)_{EE'} 
\langle E'|\hat \varphi(\vec r)|G\rangle 
\Biggr)
\label{G_ret}
\end{eqnarray}
where now $|E\rangle$ denotes excited states of the system in the 
{\em absence} of tunneling, {\em i.e.} $|E\rangle$ is a direct product of the
excited states of the isolated cluster and the excited states of the
isolated substrate and $\hat E=\delta_{E,E'}E$.  { The operator 
$\hat \varphi^\dagger(\vec r)$ ($\hat \varphi(\vec r)$) creates 
(annihilates) an electron at position $\vec r$ on the cluster.}
We average over the 
ground state multiplet $|G\rangle$ with degeneracy  $N_G=2S_0+1$, and $M^R$ 
is the generalized retarded self-energy matrix.

\begin{figure}
\centerline{\epsfig{figure=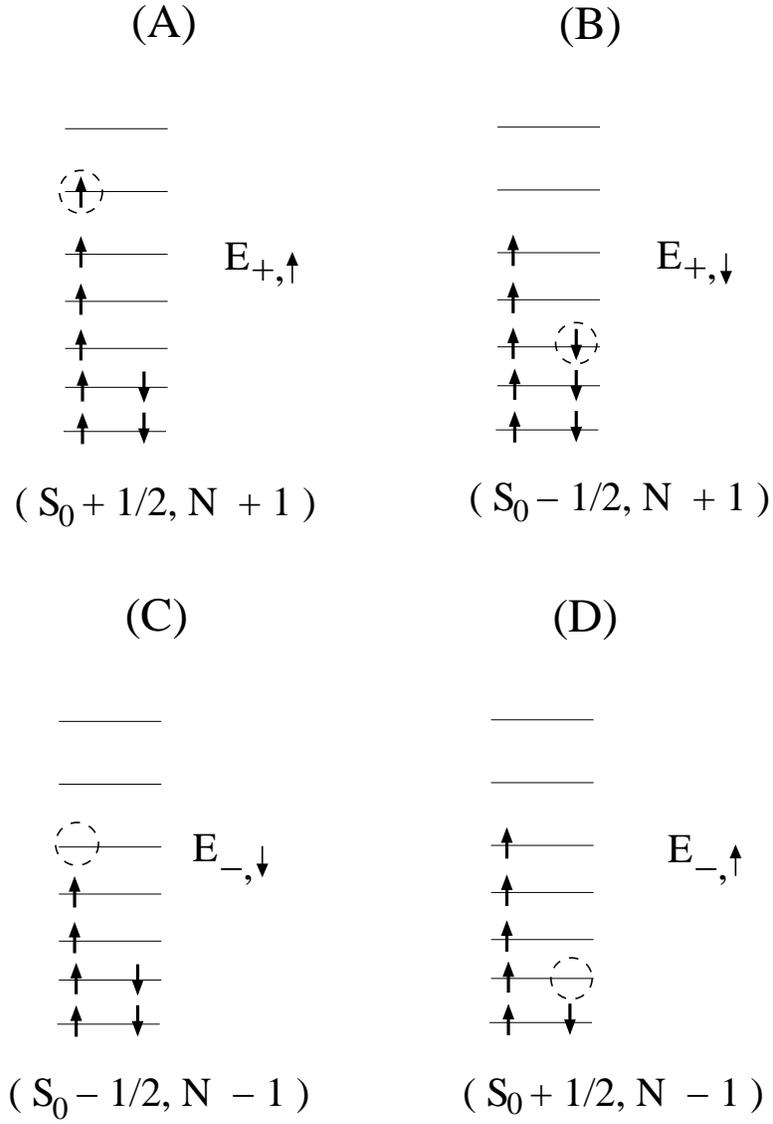, width=4 in}}
\caption{Charging excitations of a ferromagnetic particle.
Circles indicate particles added to (removed from) the ground state.
Fig.~A (B): Lowest energy state for adding
    a majority (minority) electron to the cluster.   
Fig.~C (D): Lowest energy state for adding 
    a majority (minority) hole to the cluster.}
\label{addition}
\end{figure}

Assuming that $\Gamma^j_0 \equiv 
2\pi \varrho_0 \sum_\mu |V^j_\mu|^2 < \mbox{min}\{\delta E_{\pm,\sigma},\delta_A,\delta_I\}$,  we can  
do perturbation theory in the tunneling.
In this case  the ground state $|G\rangle$ can be  approximated 
in leading order by  the product of the independent ground states 
of the cluster and the  metal, and the summation over the excited 
states $|E\rangle$  turns into a sum  over the states 
$|E_{\pm,\sigma}^j\rangle $ 
 defined in Sec.~\ref{sec:itin} (see Fig.~\ref{addition}).  
With these assumptions the retarded self-energy $M^R(\omega)$ is approximately diagonal in 
$E$ and $E'$   
\begin{equation}
M^R(\omega)_{EE'} \approx {\rm real\;\; part} 
-i \delta_{EE'}
\pi \sum_n \delta(\omega-E_n) |\langle n| \hat V|E\rangle |^2 \;,
\end{equation}
where summation is carried out over all possible 
intermediate states $|n\rangle$ { and $\hat V$ is given by Eq.~(\ref{V_use})}. The  real part of the  
self-energy  produces  only a slight  shift of the levels, and it can be
neglected.  The imaginary part, on the other hand, describes the  
broadening of the spectrum.

Within these approximations the STM spectrum can be expressed as:
\begin{equation}
{dI\over dV} \propto 
\sum_{j,\sigma,\pm} |\varphi_j(\vec r)|^2
{  \Gamma_{\pm, \sigma}^j \over (\omega \mp \delta E^j_{\pm,\sigma})^2 +
(\Gamma^j_{\pm, \sigma}/2)^2}
\label{dI_dV_final}
\end{equation}
where $\Gamma^j_{\pm, \sigma}$
is the total decay rate  of the excited state with excitation energy
$\delta E^j_{\pm,\sigma}$. 

Eq.(\ref{dI_dV_final}) assumes that no 
two  states of the cluster 	in the same spin sector 
are closer to each other than $\Gamma_0$. 
Otherwise they may produce  Fano-type  resonances  due to the 
interference between electrons (or holes) that tunnel through these states.
For our mean field models with maximum level repulsion, however,  this 
condition is always satisfied if $\Gamma_0 < \delta_I, \delta_A$.
Also, if the ground-state energy difference for the cluster 
with $N$ and $N\pm1$ electrons on it (or with spin $S_0$ and $S_0\pm1$) 
is too small compared to $\Gamma_0$
the initial state will not have a definite electron number (or spin) and 
our computation  breaks down.

\begin{figure}
\centerline{\epsfig{figure=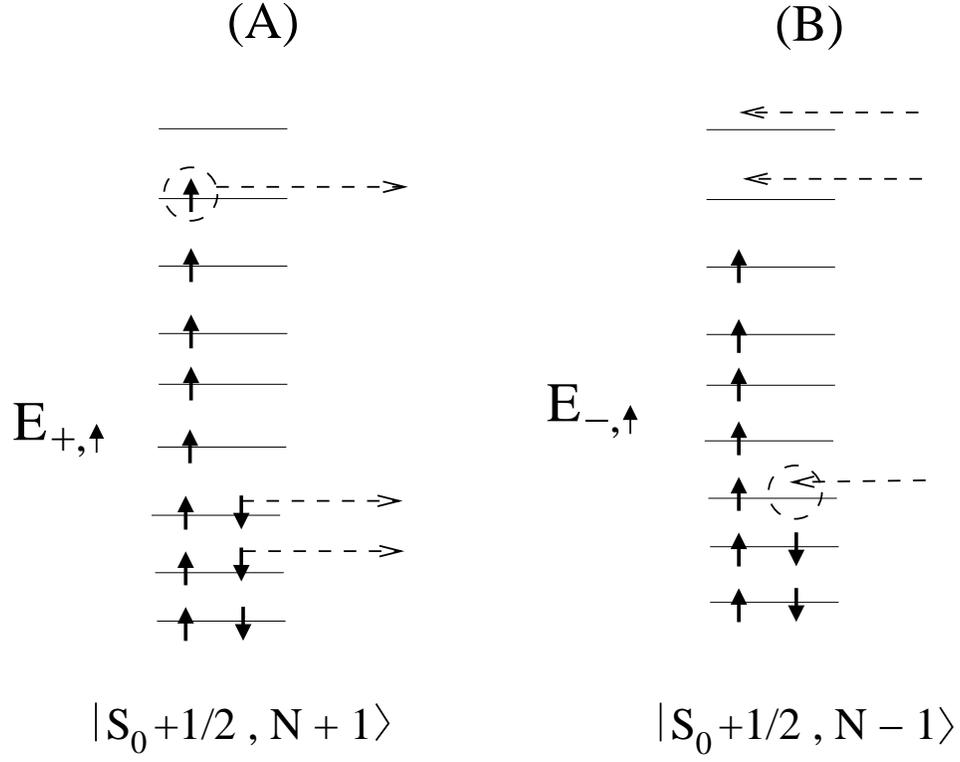, height=4 in}}
\caption{The dashed lines represent possible possible  decay channels for 
the states $|E_{+,\uparrow}\rangle$($N+1$ electrons) and 
$|E_{-,\uparrow}\rangle$ ($N-1$ electrons) to
lower energy states with $N$ electrons. In an actual decay, only one of 
processes indicated by the dashed lines would occur.} 
\label{decay}
\end{figure}

 For $\delta E^j_{\pm,\sigma} >\delta_{A,I}$ we find that the decay rate 
$\Gamma^j_{\pm, \sigma}$ is typically larger than the single 
particle tunneling rate $\sim \Gamma^j_0$. This can be understood as follows:
Assuming that the  STM-cluster voltage is already  large enough to 
overcome the Coulomb  blockade,  an energy  $\sim E_C/2$ is transfered 
to the cluster as an electron or hole hops onto the cluster
from the STM tip. The extra charge carrier then tunnels 
into the leads. However, for $\delta E^j_{\pm,\sigma} >\delta_{A,I}$ 
the island can be
left behind in a state that contains electron-hole excitations. 
{ Assuming for the sake of simplicity that  
$\Gamma^j_0 = \Gamma_0$ is independent of the 
single particle state index $j$ we obtain { the result that the width of a given level
is proportional to the number of ways it can decay:}
$\Gamma^j_{\pm, \sigma}  \approx N^{(j,\pm,\sigma)}_{\rm decay}
\Gamma_0$, where 
$N^{(j,\pm,\sigma)}_{\rm decay}$ is the number of island states with $N$
electrons whose energy $E(N)-E_G<\delta E^j_{\pm,\sigma}$ and which
are accessible
by the addition (or subtraction) of just one electron from the state
corresponding to $\delta E^j_{\pm,\sigma}$.  There is no simple expression
for the $N^{(j,\pm,\sigma)}_{\rm decay}$'s; they depend sensitively on the
level $j$, $N_A$, $E_C$ 
and the {\em fluctuating} mesoscopic parameters $d_0$ and $n_g$, 
and we have to determine them for each set of parameters 
and each level separately. 

{ Within the mean field model we are using and 
to leading order in the tunneling $\hat V$ the number of decay channels of
the state with energy $E^j_{\pm,\sigma}$ does not depend on $j$. 
This is an artifact of the mean field approximation: 
In higher orders of the tunneling (virtual tunneling processes) or in the 
presence of electron-electron interactions on the island (which may also 
change level occupations) excited states with larger energies generally
have a larger decay amplitude.}
 Examples of
decay processes for a given excited state are shown in  Fig.~\ref{decay}.}

\begin{figure}
\centerline{\epsfig{figure=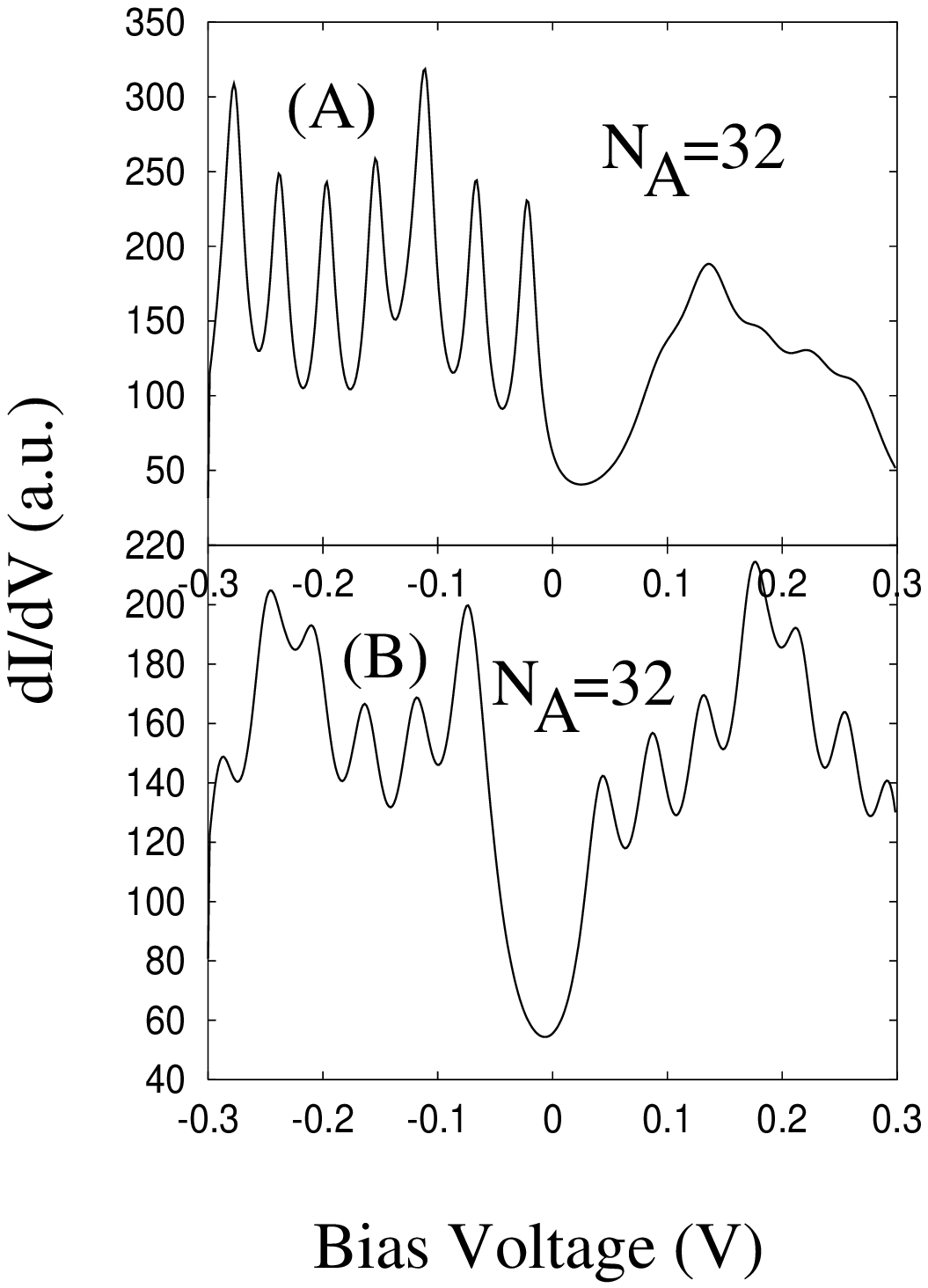, width= 5 in}}
\caption{Calculated STM spectra of nanometer-size ferromagnetic 
clusters with parameters $N_A=32$, 
$E_C^+=.01{\rm eV}$, $E_C^-=.08{\rm eV}$, $\Gamma_0=.02{\rm eV}$, and  
$\mu=0$  at 4 K (see main text).
We took  $\Gamma_0^j =2\pi|V_j|^2 \varrho_0 \approx \Gamma_0$, 
independent of $j$, and   $|\varphi({\vec r})|^2 = const$
in Eq.~(\ref{dI_dV_final}). 
In Fig.~A we set $d_0=.07{\rm eV}$ while in Fig.~B
 $d_0=-.05{\rm eV}$. The difference between the two cases
demonstrates the sensitivity to mesoscopic parameters.}
\label{spectrum}
\end{figure}

 According to Eq.~(\ref{dI_dV_final}), the weakly  coupled cluster
spectrum is simply a series of Lorentzian peaks centered at the
particle and hole addition energies determined by Eq.~(\ref{model_I}), 
Eq.~(\ref{U}) and the actual band structure of Co~\cite{papa} with a 
weight modulated by a random amplitude
$|\varphi_j(\vec r)|^2$. In our calculation (See Fig.~\ref{spectrum})
 of the STM tunneling
conductance we ignore (i) the amplitude modulation, (ii) fluctuations in the
level spacing of the cluster beyond that given by band structure 
calculations\cite{papa} (as we discussed in Sec.~\ref{sec:itin}) 
and (iii) fluctuations in the hybridization  
$\Gamma^j_0\approx \Gamma_0$. These effects could be 
included with some effort, but they would not be expected to change our
conclusions.

In principle,  contributions from phonon scattering or
electron-electron interactions also contribute to the relaxation rate 
$\Gamma^j_{\pm, \sigma}$; however, in the experiments of 
Ref.~\onlinecite{teri}
these can probably be neglected with respect to
the single particle relaxation channel considered here. 

To compute the tunneling spectrum of an actual cluster with $N_A$ atoms we
generated a discrete set of levels $\epsilon_j$ with a  
level spacing corresponding to the single particle 
density of states in Co.\cite{papa}
As discussed above, the  non-uniform density of states is 
important to obtain a predominantly antiferromagnetic coupling, 
but is also important to obtain a  stable ground state with a partially 
polarized band. 
The STM spectrum for low bias voltages is governed by 
the level structure near $\epsilon_A \mbox{ and } \epsilon_I$ since these
are the only states probed in an STM measurement at low voltage bias.  
The majority and minority level spacing at these energies can be 
estimated from band structure calculations as $\delta_A$=5.55 eV/$N_A$, 
$\delta_I$=1.43 eV/$N_A$ and $S_0 = 0.855 N_A$.\cite{papa}  
For Co, $\Delta_s=\epsilon_A-\epsilon_I \approx 2$eV. 

For the sake of simplicity we set 
$\bar\epsilon =(\epsilon_I +\epsilon_A)/2 = 0$ in our computations, effectively
absorbing it into $n_g$.
This corresponds to a specific choice of contact potential between the lead 
and the cluster, and does not influence the overall 
features of the spectrum.  

Eq.~(\ref{dI_dV_final}) assumes the cluster is in its ground state before
the electron from the STM tunnels into it.  Thus, in the
numerical calculation it is important to ensure that the 
parameters of the model give a ground state with definite $S_0$ and
$N$. Using Eq.~(\ref{model_I}) and Eq.~(\ref{U}) and keeping 
track of the $1/N_A$
corrections which are all put into $d_0$, one can derive the conditions for
stability of the ground state, which we assume to have $N_s \approx 1.71 N_A$ 
singly occupied levels. 
{ By considering  the fluctuations
in the ground state spin shown in Fig.~\ref{states}B and C one can show that 
to guarantee stability,  $d_0$ must satisfy the conditions:
\begin{equation}
-{N_A \over S_0}\delta_I < d_0 < {N_A \over (S_0+1)}\Biggl(\delta_A -{\Delta_s \over S_0}\Biggr)\;.
\label{do_stability}
\end{equation}
For Co this simplifies to $-1.67 {eV \over N_A}< d_0 <3.21 {eV \over .855 N_A +1}$.
Similar conditions
can easily be derived for the quantities 
$E_C^+\mbox{ and } E_C^-$ by considering  fluctuations to the $N\pm1$ 
manifold of states like those in 
Fig.~\ref{addition} and requiring $\delta E_{\pm,\sigma}>0$. For our
model applied to Co we obtain $E_C^+>-1.0 {eV \over N_A} \mbox{ and }
E_C^->2.48 {eV \over N_A}$.
}

A calculation of the tunneling spectrum of a ferromagnetic cluster with 
$N_A=32$ atoms and a corresponding spin $S_0= 27$
is shown in Fig.~\ref{spectrum}. 
For the computations we used the parameters 
$E_C^+=.01{\rm eV}$, $E_C^-=.08{\rm eV}$ and $\Gamma_0=.02{\rm eV}$
in both figures.  Then once $d_0$ is given all the $\delta E^j_{\pm,\sigma}$ 
are determined via relations similar to Eq.~(\ref{mu}).
The quantity $N_{\rm decay}^{(j,\pm,\sigma)}$ is computed
numerically.  In Fig.~\ref{spectrum}A  $d_0=.07{\rm eV}$, 
and the ground state is 
close to being unstable against the lowest lying 
state within the $(N,S_0+1)$ subspace, while in 
 Fig.~\ref{spectrum}B  $d_0=-0.05 {\;\rm eV}$, and the ground state is 
close to the lowest lying  state within the $(N,S_0-1)$ 
subspace.

In the  spectrum in Fig.~\ref{spectrum}A 
there is a series of sharp peaks at  negative bias (electrons
are removed from the cluster) while the spectrum at positive bias 
(electrons are added to the cluster) is rather smooth.
These features correspond very well with asymmetrical 
features seen in some spectra of Ref.~\onlinecite{teri}.  
On the other hand, the lower spectra of 
Fig.~\ref{spectrum}B has less contrast between positive and negative biases. 
The only parameter that has been changed between the two spectra is $d_0$ 
of Eq.~(\ref{U}).  { Different values of $d_0$ lead to different 
values of the
number of decay channels $N_{\rm decay}^{(j,\pm,\sigma)}$ for positive 
and negative bias, and
hence lead to different widths of the peaks in dI/dV.}
This calculation shows how mesoscopic fluctuations
may make a significant qualitative difference in the spectra of
two clusters with the same $N_A$, $E_C$ and $n_g$. For real clusters 
mesoscopic fluctuations in  $E_C$ and $n_g$ are also expected for fixed $N_A$
and these will also lead to qualitative changes in the spectra.

A careful comparison of our calculated spectra 
corresponding to the first few charging peaks with
the experimental spectra shows that at voltage biases up to $\approx \pm E_C$
our calculations are in agreement with experiment, however at larger 
biases, eV$>E_C$ (when {\em two}
additional electrons (or holes) could be added to the cluster at the same
time), the experimental spectra show additional structure, not present in
our calculations, presumably due 
to many-body and/or non-equilibrium excitations left out of our simple model.

As we will see in the next section, for the cluster size of Fig.~\ref{spectrum}
the estimated Kondo temperature 
is significantly below the experimental temperature, 
and no Kondo peak appears at zero bias.

\subsection{Estimating the Kondo Temperature}
\label{sec:estimate_kondo_temp}
It is obvious that Eq.~(\ref{dI_dV_final}) does not include Kondo
correlations and therefore does not produce a Kondo peak in the
spectrum.  However, it is generally true that when the Kondo effect is
present, it produces a peak of width $\sim T_K$ at the Fermi level
with relative weight $\sim { T_K\over \Gamma_0}$ due to the approximate
unitary scattering, provided the temperature is 
less than $T_K$, $T< T_K$.\cite{hewson,nozieres}  

In order to estimate $T_K$ and its dependence on cluster size, we will
use the results of Sec.~\ref{onepoint} for a cluster contacting the
substrate in only a single point of contact.  There are several reasons 
why we believe this may be a reasonable assumption to make for 
Co nanoclusters on metallic nanotubes.\cite{teri}  
First of all, the ratio of broadening $\Gamma$ of energy 
levels on the cluster compared to their separation $\sim \delta_{A,I}$
is experimentally found to be almost independent of the cluster size. 
Since  $\delta_{A,I}\sim 1/N_A$ 
and $\Gamma \sim N_P/N_A$ ($N_P$ is the number of points of contact), 
this indicates that the number of effective tunneling 
points is approximately  constant for the clusters investigated 
in the measurements.  In fact, since the diameter of the clusters (.5-1 nm) of
Odom {\em et al.}~\onlinecite{teri} 
are comparable to the diameter of
a nanotube there is appreciable curvature at the cluster-nanotube interface, 
and it is quite possible that the cluster touches the surface of the 
nanotube only at a few points. Since the tunneling amplitude is 
exponentially  sensitive to the tunneling distance, 
it is also possible that only one or two of these points 
dominates the conductance between the cluster and the nanotube, 
resulting effectively in a coupling in the form of
a single point of contact.

The standard expression for the Kondo temperature in the Kondo model 
is given by\cite{hewson}
\begin{equation}
T_K= D \sqrt{J^{\rm eff}\varrho_0} e^{-1/ J^{\rm eff}\varrho_0}
\end{equation}
where $D$ is the bandwidth of the conduction electrons of the metallic host 
and $\varrho_0$ is the density of states of the host at the impurity site.
For a ferromagnetic cluster, however, this formula is incorrect, 
since the Kondo Hamiltonian provides an  appropriate description of 
the cluster dynamics only below the characteristic energy of 
inelastic excitations, $E_{\rm inel}\equiv {\rm min}\{E_C,\delta_{A,I} \}$. 
As shown for example in the two-level system model,\cite{altshuler}
above this energy scale fluctuations to  various excited states 
destroy the coherent spin processes leading to the Kondo 
effect, and in the renormalization group approach in the regime
above $E_{\rm inel}$ the Kondo coupling remains 
unrenormalized. Therefore, for a single point of contact we 
estimate the Kondo  temperature as 
\begin{equation}
T_K\sim \mbox{\rm min}\{E_C,\delta_{A,I} \}\sqrt{J^{\rm eff}\varrho_0} 
e^{-1/ J^{\rm eff}\varrho_0},
\label{eq:T_K}  
\end{equation}
where we replaced the bandwidth $D$ of the conduction 
electrons by the energy gap for inelastic processes on the 
cluster, $D \to {\rm min}\{E_C,\delta_{A,I} \}$.
Eq.~(\ref{eq:T_K}) is only of logarithmic accuracy. 
In general, $T_K$ contains  an overall prefactor  
that incorporates corrections from  higher excited states  
as well as  possible mesoscopic fluctuation effects. 
This prefactor is usually of the order of unity, 
however in some cases  it can be quite large and considerably  
increase the Kondo temperature.\cite{cfield} 

In principle, the quantities $E_C\mbox{ and } \delta_{A,I}$  can be
obtained directly from the STM spectra, and 
$J^{\rm eff}\varrho_0$ can also be related to the spectra via
the width of the energy levels of the cluster.  
Recalling $\Gamma^j_0=2\pi |V^j|^2\varrho_0$ 
we can write
\begin{equation}
J^{\rm eff} \varrho_0 \approx {\Gamma_0 \over 2 \pi S_0}
\Biggl[P \int_{-\infty}^\infty {\Delta_s \varrho(\xi) d\xi \over (\xi -\epsilon_I)
(\epsilon_A -\xi)}
-\varrho(\epsilon_I) \mbox{ln} \Biggl({\delta E_{+,\downarrow}  \over \delta E_{-,\uparrow}} \Biggr)
+\varrho(\epsilon_A) \mbox{ln} \Biggl({\delta E_{+,\uparrow}  \over \delta E_{-,\downarrow}} \Biggr)\Biggr]\;,
\label{eq:J_rho}
\end{equation}
where we have used Eq.~(\ref{eq:Jestimate}) and taken $|V^j|^2\to \langle 
|V_j|^2\rangle$.  Recall that $\varrho(\epsilon) \propto N_A$.

%\begin{table}
%\begin{tabular}{lcccccc}
%\hline \hline
%   $N_A$   & $\Delta_s$ &   $\Gamma$   &   $\delta_I$   &   $E_C$  
%& \phantom{n}$T_K^{\rm exp}$ &  \phantom{n}[$T_K^{\rm est}(C_K = 500)$] \\
%    $8^*$ & 2 eV & $.24^*$ eV & .24 eV & $.36^*$ eV 
%& $\sim80^*$ K & [80 K]\\
%   16& 2 eV & .12 eV & .12 eV & .23 eV & --- &  [$10^{-4}$ K]\\
%   $32^*$ &2 eV & $.06^*$ eV & $.06^*$ eV & $.10^*$ eV &  --- 
%& [$10^{-12}$ K]
%\\ \hline \hline
%\end{tabular}
%\caption{\label{kondo_temps} $T_K$ for Different Cobalt Cluster Sizes.  
%The numbers with an asterisk ($^*$) were taken from the experiments of 
%Ref.~\onlinecite{teri}.  
%The remaining numbers were filled in based on our estimates.  The estimated 
%Kondo temperatures, $T_K^{\rm est}$, were estimated using 
%Eq.~\ref{eq:T_K} and Eq.~\ref{eq:J_rho} 
%with $C_K=500$ as described in the text.  The Kondo temperature
%is suppressed quickly with increasing cluster size.
%The $\delta_I$ for $N_A=32$ is consistent with what one expects for 
%Co.\cite{papa} The scaling of $\Gamma$ with $N_A$ for $N_A=8$ and $N_A=32$
%is consistent with the Co clusters making electrical contact in only a 
%single point. For $N_A=8$ mesoscopic fluctuations in Eq.~(\ref{eq:Jestimate})
%of the order of 10\% can lead to fluctuations in $T_K$ of nearly two orders
%of magnitude.}  
%\end{table}

As is evident from Eq.~(\ref{dI_dV_final}), the actual level
width of cluster excited states observed in experiment is 
$\Gamma^j=N_{\rm decay}\Gamma^j_0$, where $N_{\rm decay}$ is the
number of energetically allowed decays for a particular excited
state in the $N \pm 1$ manifold as described in Sec.~\ref{connection}A.
According to our model STM spectra calculations (Sec.~\ref{connection}A) 
(which computes $N_{\rm decay}^{(j,\pm,\sigma)}$ numerically)
$N_{\rm decay}$ is typically 1-3 for clusters with 7-30 atoms.
As the most favorable case for obtaining an experimentally consistent
$T_K$, we take $N_{\rm decay}$=1 in our estimates.

To obtain an  estimate of the Kondo temperature 
 for Co atoms adsorbed on nanotubes 
 we used the experimental data of Ref.~\onlinecite{teri}.  
For a cluster experimentally estimated to have $N_A = 8$ atoms, the value 
$\Gamma \sim 0.24 {\rm eV}$ and $E_C \sim .36 {\rm eV}$ can be 
directly determined from the 
experimental STM spectrum of cluster.
Unfortunately, the level spacing $\delta_{I,A}$ cannot be 
determined directly from the STM spectra for these small clusters, 
but can we  estimate them by rescaling the level spacing measured 
at larger clusters, giving $\delta_I \sim  0.24$ eV. 

Assuming that the spin splitting  
takes its bulk value, $\Delta_s \sim 2 {\rm eV}$ we can 
then estimate $T_K$ using Eqs.~(\ref{eq:J_rho}) and 
(\ref{eq:T_K}). We numerically evaluated  both the integral in 
Eq.~(\ref{eq:Jestimate}) as well as the discrete sums in 
Eqs.~(\ref{J_d}-\ref{J_s}) using the actual Co density of states\cite{papa}
(assuming a single point of electrical contact).
Neglecting the mesoscopic fluctuations in $\delta E_{\pm,\sigma}$, the
integral and the discrete sum were found to be within 10\% of each other
for $N_A=$8, 16 and 32. The Kondo temperature estimated this 
way for $N_A=8$, $T_K\sim 0.16$ K  turned out to be about a factor 500 
smaller than the experimentally observed Kondo temperature, 
$T_K^{\rm exp} \sim 80$ K. { (We obtain a value of $J^{\rm eff}
\varrho_0 = .12$ for $N_A=$8.  This would need to be increased by a factor of 
$\sim$ 2.5 to reach agreement with experiment.)}   Mesoscopic 
fluctuations of $E_{\pm,\sigma}$ may increase (or decrease) $J^{\rm eff}$ 
by $\sim 10 \%$.   Furthermore, our integrals
(and sums) were evaluated with the assumption that $V_j$ is independent of
energy.  This is not strictly true and will lead to additional fluctuations
in $J^{\rm eff}$. These mesoscopic fluctuation effects typically change the 
value of $J^{\rm eff}$ altogether by $10-15 \%$.  It seems unlikely 
such fluctuations could bring
the theoretical estimate of $T_K$ into the experimentally observed range.
%However, this would mean that relatively few clusters of $N_A\sim8$ Co atoms
%should exhibit a Kondo effect at 5 K because their $T_K$ would be too small. 
%Experimentally, a few clusters with $N_A\sim8$ showed a Kondo effect and a few
%did not suggesting both that (i) fluctuations are important in determining 
%the $T_K$ of a cluster and (ii) that the clusters with an observable 
%Kondo temperature are not as rare as they should be according to our
%theoretical estimates which require relatively large fluctuations to agree
%with the experimental $T_K$. 
Thus, there is a discrepancy between 
experiment and theory.  There may be several explanations   for this 
disagreement that  
we discuss in detail in Sec.~\ref{discussion} of the paper.

Eq.~(\ref{eq:J_rho}), on the other hand, is in qualitative 
agreement with the experiments in that it predicts an extremely 
 rapid decrease of $T_K$ with increasing cluster size. 
Since for a single point of contact $\Gamma_0 \sim 1/N_A$, 
$S_0 \sim N_A$, and $\varrho(\xi) \sim N_A$, the dimensionless
exchange coupling scales as $J^{\rm eff} \varrho_0 \sim 1/N_A$. 
We also verified numerically that the scaling $J^{\rm eff} \propto 
{1\over N_A}$
of Eq.~(\ref{eq:Jestimate}) is maintained for the discreet sum in  
Eqs.~(\ref{J_e}-\ref{J_s}).  

{ Note that if one were to assume multiple contacts between the 
cluster and substrate, while keeping the decay rate fixed, then one
would obtain smaller values for the typical tunneling matrix
element $V_\mu^{j,k}$
and, hence, a lower Kondo temperature.}

{ As we found in Sec.~\ref{onepoint}, the dimensionless coupling
constant for a non-magnetic cluster, $J_{1/2}^{\rm eff} \varrho_0 \sim 1/N_A^{1-\alpha}$, so that the Kondo temperature is suppressed less rapidly with increasing
$N_A$.  In this sense, the large spin of a 
ferromagnetic cluster does not ``help'' the Kondo effect in anyway.}

\section{Local Moment Clusters}
\label{localmoment}
\subsection{Local moment mean field model}

In many magnetic materials it is more appropriate to think of 
a localized d- or f-level (as in the Anderson Model)
than to think of strongly  hybridized 
s, p and d bands.  These local moments may couple to each-other 
ferromagnetically and  produce ferromagnetism. 

For small enough magnetic clusters with large enough Curie 
temperature,  at low enough temperature the local moments 
form a large and rigid ferromagnetic spin, $S_d$, that couples 
to the extended states (a `conduction band') of s- and p-character. 
The simplest Hamiltonian that one can conceive 
to describe this situation reads
\begin{equation}
\hat H_{\rm cluster}=
\sum_{j,\sigma}\epsilon_j c^\dag_{j \sigma}c_{j \sigma} +{J\over
  N_A}\hat S_c \cdot \hat S_d + {E_C\over 2}(\hat
  N -n_g)^2, 
\label{model_II_final}
\end{equation}
with $\hat S_c$ and $\hat S_d$ the total spin of the extended states and 
the local moments, respectively.
A justification  of Eq.~(\ref{model_II_final})  is given 
in Appendix~\ref{sdderivation}. Similar to the itinerant model, 
the first term describes the kinetic  energy of the extended  electron states, 
and the third term accounts for the finite charging energy of the 
cluster.  The second term of Eq.~(\ref{model_II_final}) describes the 
exchange interaction between the local moments and the extended states
and tends to polarize the latter.
We assume in what follows that the total spin of the conduction
electrons is much smaller than that of the localized electrons. The
exchange $J$ is typically antiferromagnetic, $J>0$, 
and the conduction electrons are polarized 
opposite to  d-electrons.\cite{himpsel}  

The local moment model has  
$S_T,S_T^z$ (the total spin and its z-component), $\{n_j\}$, 
$S_d$, and $S_c$ as conserved quantum numbers, and can thus 
be  diagonalized exactly. 
The ground state  is given by
\begin{equation}
|S_T,S_T^z\rangle_{S_c,S_d}^{N}=\sum_{S_c^z,S_d^z;S_c^z+S_d^z=S_T^z}\langle
 S_c,S_d;S_c^z,S_d^z|S_T,S_T^z\rangle
|S_c,S_d;S_c^z,S_d^z\rangle^{N}\;,
\label{model_II_states}
\end{equation}
where $|S_c,S_d;S_c^z,S_d^z\rangle^{N}=|S_d,S_d^z\rangle|S_c,S_c^z\rangle^{N}$ and $S_T=S_d-S_c$ if $J>0$. The state $|S_c,S_c\rangle^{N}$ can be 
computed from Eq.~(\ref{ground_state}) with $S_c$ replacing $S_0$ everywhere,
 and $|S_c,S_c^z\rangle^{N}$ is computed
from Eq.~(\ref{spin_lower}) with $S_c$ replacing $S_0$ and $S_c^z$
replacing $S^z$. 

Stability of the ground state implies the relation 
\begin{equation}
J={N_A \over (S_d + 1)}\Delta_s + O(1/N_A)\;,
\label{model_II_stability}
\end{equation} 
where $\Delta_s$ is now the band splitting of the conduction band.  

Considering the type of particle-hole excitations shown in Fig.~\ref{states}
and using Eq.~(\ref{model_II_stability}), one finds
that the excitation  spectrum is very similar to that of the itinerant model.
In particular, we find that for charge fluctuations 
$\delta E_{\pm,\sigma}\sim E_C/2$,
and spin fluctuations have a gap  $\sim \delta_A, \delta_I$.

\subsection{Computation of the exchange coupling}
\label{model_II_coupling}
The coupling constants $\tilde J^{\mu \nu}$ of
the local moment model depend on the sign of the exchange coupling, $J$, of
Eq.~(\ref{model_II_final}). Let us first focus on the case $J>0$ and 
$S_d > S_c$.
 For the local moment model, $\langle f|\hat H_{\rm Kondo}^{\rm eff}|i\rangle={1\over 2}\sqrt{2S_T} \tilde J^{\mu  \nu}$.
To evaluate the RHS of Eq.~(\ref{pert})  we must again evaluate contributions
from the conduction electron states with double, single and no occupation:
\begin{equation}
\tilde J^{\mu\nu} = \tilde J^{\mu\nu}_{d} +\tilde J^{\mu\nu}_{s} + \tilde J^{\mu\nu}_{e}\;.
\end{equation}
For the sake of simplicity, let us consider the contribution
 of singly occupied  levels, $\tilde J^{\mu\nu}_s$.
Various matrix  elements of the type 
${}_j^{N+1}\langle S_T+1/2,S_T-1/2|c_{j \downarrow}^\dagger|S_T,S_T\rangle^N$
arise in course of the evaluation of $\tilde J^{\mu\nu}$, and 
in contrast to the itinerant cluster model, the intermediate 
states of the local moment model
 (with $J>0$ and $S_d> S_c$) have an {\it increase} in total spin on the 
cluster in this case. To evaluate 
${}_j^{N+1}\langle S_T+1/2,S_T-1/2|c_{j \downarrow}^\dagger|S_T,S_T\rangle^N$
we first expand  $|S_T,S_T\rangle_{S_c,S_d}^N$ and 
$|S_T+1/2,S_T-1/2\rangle_{S_c-1/2,S_d}^{N+1}$ using 
Eq.~(\ref{model_II_states}) to obtain  
\begin{eqnarray}
{}_j^{N+1}\langle S_T+1/2,S_T-1/2|c_{j \downarrow}^\dagger|S_T,S_T\rangle^N
= \nonumber \\
\sum_{S_c^z=-S_c}^{S_c}
\langle S_c,S_d;S_c^z,S_T-S_c^z|&&S_T,S_T\rangle \nonumber \\
\langle S_c-1/2,S_d;S_c^z-1/2,S_T-S_c^z|&&S_T+1/2,S_T-1/2\rangle
\nonumber \\
{}_j^{N+1}\langle
S_c-1/2,S_c^z-1/2|&&c_{j \downarrow}^\dagger|S_c,S_c^z\rangle^N. 
\label{matrix_element}
\end{eqnarray}
What remains to be computed in Eq.~(\ref{matrix_element}) is the matrix element
${}_j^{N+1}\langle S_c-1/2,S_c^z-1/2|c_{j \downarrow}^\dagger|S_c,S_c^z\rangle^N$.
To determine this, we use the states of Eq.~(\ref{spin_lower}) with $S_c$
replacing $S_0$.  The overlap is computed by first directly evaluating
${}_j^{N+1}\langle S_c-1/2,S_c-1/2|c_{j \downarrow}^\dagger|S_c,S_c\rangle^N$
and then applying the Wigner-Eckhart Theorem for general $S_c^z$. This yields
${}_j^{N+1}\langle
S_c-1/2,S_c^z-1/2|c_{j \downarrow}^\dagger|S_c,S_c^z\rangle^N =  
\sqrt{{S_c +S_c^z\over 2 S_c}}$.  This can then be substituted into
Eq.~\ref{matrix_element} which finally gives
\begin{eqnarray}
&& {}_j^{N+1}\langle S_T+1/2,S_T-1/2|c_{j \downarrow}^\dagger|S_T,S_T\rangle^N
 \nonumber \\
&&\phantom{mmm}=\sum_{S_c^z=-S_c}^{S_c} \sqrt{{S_c +S_c^z\over 2 S_c}}
\langle S_c,S_d;S_c^z,S_T-S_c^z|S_T,S_T\rangle \nonumber \\
&&\phantom{mmm} 
\langle S_c-1/2,S_d;S_c^z-1/2,S_T-S_c^z|S_T+1/2,S_T-1/2\rangle
\label{m_1}
\end{eqnarray}
The results of the evaluation of all the matrix elements on the right 
hand side of Eq.~(\ref{pert}) for the local moment model as well as the 
expression equivalent to Eq.~(\ref{eq:J_sum}) are somewhat lengthy 
so we relegated them to Appendix~\ref{matrix_model_II}.
We evaluated them numerically for $S_T=S_d-S_c$
(antiferromagnetic $J$ and $S_d>S_c$) and  found that the 
single particle contribution to the 
final exchange coupling differs in an overall  sign from the itinerant
model result of Sec.~\ref{perturbativelycoupled}, and it is 
ferromagnetic.  On the other hand,
if $J$ is ferromagnetic, then $S_T=S_d+S_c$ and the matrix
elements above agree in sign with those of 
Sec.~\ref{perturbativelycoupled}.

This can qualitatively be understood as follows: 
Suppose the total spin, $S_T$, of the cluster points upward. Then  
  by assumption  $S_d> S_c$ the spin of the 
d-levels, $S_d$, also points up. If the internal interaction, $J$,  between 
$S_d$ and $S_c$ is antiferromagnetic, delocalized electrons of 
the cluster with spin down will partially screen the local spin, $S_d$, 
so that the singly occupied states tend to have have spin down. 
A substrate conduction electron that hops on a singly occupied state
must have, therefore, spin up that is parallel to the total spin, resulting
in a ferromagnetic contribution to the effective interaction 
between  the total cluster spin and the  substrate. 
On the other hand,  due to the
antiferromagnetic interaction with the local spin, hopping 
to  empty states
with spin down have an energy smaller than those with spin up (parallel to the
total spin), and give rise to an antiferromagnetic contribution 
to the effective interaction between the cluster spin and the substrate.
The case $J<0$ can be understood along the same lines.

%Conduction electrons in the metal ``communicate'' directly 
%with only extended states on the cluster through the tunneling 
%process. 
%Tunneling  generates therefore an antiferromagnetic 
%coupling (in second order perturbation theory)  between the electrons 
%in the metal and the conduction 
%electrons on the cluster. If the coupling between 
%$S_c$ and $S_d$ is also antiferromagnetic and $S_d>S_c$,
%this results in an effective {\em ferromagnetic coupling} of the conduction
%electrons' spin to the total spin of the cluster.  

Similar to the itinerant case, the signs of  $\tilde J^{\mu\nu}_{d}$ 
and $\tilde J^{\mu\nu}_{e}$ are always opposite 
to that of $\tilde J^{\mu\nu}_{s}$ regardless of the sign of $J$.
Therefore, there is in general a competition between these terms, 
and the sign of the final coupling depends on specific 
band structure features.

For an antiferromagnetic coupling, $J>0$, in the limit where 
$\Delta_s, E_C \gg \delta$, and $S_c, S_d, S_T \gg 1$
we can obtain the  following simple estimate for a single point of contact, 
analogous to Eq.~(\ref{eq:J_rho}):
\begin{equation}
\tilde J^{\rm eff} \varrho_0 \sim -
{\langle|V_j|^2\rangle \over S_T }
\Biggl[P \int_{-\infty}^\infty {\Delta_s \varrho(\xi) d\xi \over 
(\xi -\epsilon_I)
(\epsilon_A -\xi)}
-\varrho(\epsilon_I) \mbox{ln} \Biggl({\delta E_{+,\uparrow}  \over \delta E_{-,\downarrow}} \Biggr)
+\varrho(\epsilon_A) \mbox{ln} \Biggl({\delta E_{+,\downarrow}  \over \delta E_{-,\uparrow}} \Biggr)\Biggr]\;.
\label{eq:Jtilde_rho}
\end{equation}
The sign of the effective coupling depends on the sign of $S_c-S_d$ (here
given for $S_c-S_d<0$).
For completeness, we also give the expression for $J<0$, which does not
depend on the relative size of $S_c$ and $S_d$ (note changes in
mesoscopic fluctuations and overall sign):
\begin{equation}
\tilde J^{\rm eff} \varrho_0 \sim 
{\langle|V_j|^2\rangle \over S_T }
\Biggl[P \int_{-\infty}^\infty {\Delta_s \varrho(\xi) d\xi \over 
(\xi -\epsilon_I)
(\epsilon_A -\xi)}
-\varrho(\epsilon_I) \mbox{ln} \Biggl({\delta E_{+,\downarrow}  \over \delta E_{-,\uparrow}} \Biggr)
+\varrho(\epsilon_A) \mbox{ln} \Biggl({\delta E_{+,\uparrow}  \over \delta E_{-,\downarrow}} \Biggr)\Biggr]\;.
\end{equation}
{ These expressions are particularly interesting:  
However large the constituent spins ($S_c,S_d$) of the cluster are, 
the cluster may still
have a large effective coupling  ${\tilde J}^{\rm eff}$ 
if the coupling $J$ between 
$S_d$ and $S_c$ is antiferromagnetic ($J>0$) and the total spin is 
sufficiently small. 
In most cases $S_c < S_d$, therefore ${\tilde J}^{\rm eff}$
is ferromagnetic and no Kondo effect develops. For 
$S_c > S_d$, however,  ${\tilde J}^{\rm eff}$ changes sign (still assuming
$J>0$) and
becomes antiferromagnetic. In this case 
a Kondo effect  occurs with an effective coupling 
proportional to $\sim 1/S_T$.
}

{ It should be pointed out that the results described above are valid
only in the weak tunneling limit.  In the strong tunneling limit, the 
relative signs of $\tilde J^{\rm eff}$ and $J$ are switched.  For example, when
the ``internal'' interaction $J>0$ and the cluster is in the {\em strong} 
tunneling regime one cannot distinguish the cluster wave functions from 
those of the host.  Thus, the $\tilde J^{\rm eff}$ would have the same
sign as $J$.  Therefore we expect in this case  $J_{\rm eff}$ 
to change sign as one gradually increases the tunneling between the 
cluster and the substrate.
We have not studied in detail how this transition would occur.}

\section{Discussion}
\label{discussion}
This work grew out of an effort to better understand the experiments of 
Odom {\it et al.}\cite{teri} in which Kondo effect was observed at low
temperatures $\sim$5 K for 
sub-nanometer Co particles adsorbed on metallic carbon nanotubes.  Our model
correctly predicts a Kondo temperature that decreases quickly with 
increasing cluster size, however our numerical estimates of $T_K$ tend to
be too small by a factor of $\sim$500 for a cluster with $N_A$=8.  

There may be several explanations for our low estimate of $T_K$. \\
(a) Mesoscopic fluctuations in Eq.~(\ref{eq:Jestimate})
can eventually increase the effective Kondo coupling and thus 
bring $T_K$ close to its experimental value. However, since the 
sign of the mesoscopic fluctuations is random, this interpretation 
would appear to contradict the experiments in which
a significant fraction of small Co clusters produced a Kondo  effect. 

{
In addition to fluctuations in  the various 
charging energies, the tunneling parameters
$V_j$ and the level positions $\epsilon_j$ also fluctuate from 
cluster to cluster. These additional fluctuations were 
neglected in Eq.~(\ref{eq:Jestimate}), since their 
contributions decrease with increasing cluster size. 
For small clusters, however, they may produce important additional 
fluctuations in $J^{\rm eff}$.
}
\\
%(b) The most likely explanation is that the spin of the cluster
%is less than the one estimated based on bulk magnetization values
%in Co. It is possible that for these small clusters
%$\Delta_s$ is for some reason reduced with respect to its bulk value. 
%In fact, a fifty percent reduction of $\Delta_s$ would be able to 
%account for the experimentally observed Kondo temperature. 
%Another possibility is that for small clusters some of the 
%spins are antiferromagnetically aligned with the 
%total spin of the cluster. Antiferromagnetic alignment
%with  25 percent of the spins could also explain the discrepancy
%between theory and experiment. \\
(b) It appears furthermore that the experiments were performed close to 
the mixed valence regime as the width of the levels is comparable to
the Coulomb charging gap. The effective Kondo Hamiltonian
we derived in second order perturbation theory may not adequately predict
$T_K$ in that case. In general, approaching the mixed valence 
regime the  Kondo temperature becomes larger than expected by the 
naive Kondo model calculation, the Coulomb gap 
shrinks, and the Kondo resonance gradually merges with the 
high-energy part of the spectrum. \\
(c) It could be possible that some of the Co atoms in the 
cluster are not strongly attached to the others, and in the 
STM  spectrum one observes the signal of these individual atoms. 
This explanation is, however, very unlikely in our opinion, 
because the Co atoms show a very strong tendency to cluster
formation, and moreover the Kondo resonance is observed
rather uniformly over the surface of clusters which are supported on 
nanotubes. \\
(d) { In our analysis, we assumed that
the anisotropy energy is smaller than the Kondo temperature, and
therefore neglected it. However, it is conceivable that
very small clusters have a considerably larger anisotropy than
our estimates based on  experiments on large clusters.\cite{ralph}
According to the STM measurements,\cite{teri}  
Co  clusters in the nanotube experiments tend to have  'pancake-shape'
and the relative position of the Co atoms is probably strongly
modified with respect  to the bulk due to the presence of the
substrate. Although the value of spin-orbit interaction on
Cobalt is not particularly large, it is still possible, that the  
highly anisotropic shape of the cluster and the deformed bonds
generate an anisotropy, that is larger or comparable to the observed
Kondo temperature, $T_K\sim 70K$.
The effect of anisotropy on the behavior of the grain is
rather complex, and  we shall discuss it in a subsequent
publication \cite{greg_next}. We would like to mention, however,
two important results that may be relevant to the experiments.
Large spin anisotropy is usually unfavorable to the Kondo
effect: In most cases it leads to an Ising-like
behavior with exponentially suppressed effective Kondo
couplings, and gives rise to a dramatic decrease of $T_K$. However,
for  very small  grains with a half-integer total spin and an almost
perfect {\em planar} anisotropy,  it can result in an
effective strongly anisotropic Kondo coupling that is considerably
{\em larger} than  the couplings in Eq.~(\ref{eq:J_rho}).
We find that for the  smallest  grains in  Ref.~\onlinecite{teri}
strong planar anisotropy  could give rise to  a $T_K$  
in the experimental range. }\\
{ (e) Another possible source of error is our assumption that 
the calculated bulk 
density of states can be used for a small cluster.  If the peak in the density 
of singly occupied states is shifted significantly from the energy value 
shown in Ref.~\onlinecite{papa}, the value of $J^{\rm eff}$ might be 
increased.  It is also possible that 
many-body corrections, omitted from our mean field model, could increase the 
value of $J^{\rm eff}$ sufficiently to account for the discrepancy with the 
experimental $T_K$.}

It is interesting to compare the results for ferromagnetic Co clusters 
with nonmagnetic Ag clusters studied on single wall 
metallic nanotubes.\cite{teri}   
The Ag clusters showed 
no Coulomb gap or discrete level spacing in the STM spectrum. This suggests 
that the Ag clusters were not in the weak tunneling regime where valence 
fluctuations can be ignored, and where an effective Kondo Hamiltonian can be 
derived for a particle with an odd number of electrons.  (Indeed, if the 
coupling to the substrate is sufficiently large, the mean number electrons on 
the cluster may be far from an integer, and the distinction between even and 
odd becomes meaningless.)  Our analysis suggests that a Kondo effect can occur 
for a particle of a non-magnetic metal, with odd electron number, if the 
coupling to the nanotube is in an appropriate intermediate regime.

It is also interesting to compare the results for Co particles on nanotubes
with measurements of several Co particles on a highly-oriented pyrolytic 
graphite (HOPG) sheet reported in Ref.~\onlinecite{teri}. The STM 
measurements did not show apparent single particle levels in the
latter case.  Assuming that the coupling to a nanotube and graphite were
not too different, this could be explained by a higher density of states 
on the HOPG surface.\cite{wallace}
(Recall that for 1 nm Co clusters on nanotubes, the
level broadening was roughly equal to the level spacing. Thus the levels may
be broadened beyond resolution on the HOPG surface.)
The STM measurements typically show a minimum in the $dI/dV$
spectrum near zero bias, when tunneling into the Co cluster on HOPG, but 
the width of
the feature is relatively large.  When fit to a Fano formula for a  Kondo
resonance, the authors of Ref.~\onlinecite{teri} obtained values of 
$T_K$ of order 700 K even for
clusters as large as 1 nm in diameter.  Since it was not possible to raise the
temperature enough to observe a temperature effect on the tunneling feature,
however, supporting evidence for existence of a
Kondo effect could not be obtained from this source.  
We note that STM measurements for tunneling directly into the
HOPG substrate also show a mininum at zero bias.

Differences in the coupling of Co clusters to a nanotube or graphite surface
may also play a role in the observed spectral differences.
Theoretical and experimental studies of STM images of graphite surfaces 
have indicated that there is an asymmetry in the local density of states
at nearest neighbor atoms.\cite{tomanek} This difference may also play a role 
in the interpretation of the spectra of Co on HOPG. Finally, it is 
possible that the matrix element for 
coupling between the cluster and the nanotube is reduced relative to the 
coupling to graphite due to the curvature of the nanotube.

\section{Conclusions}
\label{conclusions}
In this paper we have studied electron scattering from ferromagnetic clusters
on a metallic substrate.  We studied  two cluster models.
The first model describes   itinerant ferromagnetism 
\cite{canali} and is  probably appropriate for the description 
of experiments such as those of Odom {\it et al.}\cite{teri} on 
Co clusters. We also proposed  another solvable cluster model, 
where spins on the d-levels are treated as {\em localized} entities. 
This latter model may be more  appropriate for 
nanoscale rare earth ferromagnets or
semiconducting ferromagnets such as GaMnAs, though in both cases 
spin-orbit interaction plays an important role and leads to 
strong spin-anisotropy effects.
 
We derived a  general expression for the Kondo couplings 
$J^{\mu \nu}$  for both  ferromagnetic cluster models.
The sign of the obtained coupling depends in both models 
on the details  of the band structure. For the itinerant model, 
virtual tunneling onto the singly occupied levels 
on the cluster induces an antiferromagnetic exchange interaction,
while doubly occupied and unoccupied levels generate 
a ferromagnetic contribution to the exchange coupling. 

We have shown that for Co clusters  the itinerant model leads to 
{\em dominantly antiferromagnetic coupling} between the cluster spin 
and the conduction electron spins. However, fluctuations to 
doubly occupied and empty states give a large ferromagnetic
contribution to the exchange coupling that reduce it to roughly half its 
original value, and thus cannot be neglected. { (As we discussed
in Sec.~\ref{perturbativelycoupled}, for the non-ferromagnetic 
spin $S=1$ islands studied in  Ref.~\onlinecite{Glazmanpaper} these 
ferromagnetic  contributions are small.)}

The exchange coupling $J^{\mu\nu}$ involves various scattering channels. 
Therefore, in principle, the cluster could produce 
a series of Kondo effects where  the spin of the cluster is 
gradually screened.\cite{leo_2} 

It is important to emphasize that in the regime of weak electron tunneling
between the metallic substrate and the cluster,  
ferromagnetism has no special role in producing the Kondo effect 
{ as we have emphasized in Sec.~\ref{sec:estimate_kondo_temp}}. 
In fact, our calculation shows that ferromagnetism  tends to {\em suppress} 
the Kondo { temperature with increasing $N_A$ more so than for the
case of a non-ferromagnetic cluster}.
{ Besides the ``strength'' of the ferromagnetism, $U$,} the Kondo scale 
is also affected by the { density of states on the cluster} and finite 
charging energy, as well as the cluster-metal conductance.  

The weak tunneling analysis we performed is only appropriate if 
the conductance between the cluster and the metal lead is smaller than the 
quantum conductance. Increasing the number of tunneling points leads to an 
increase in  the cluster-metal conductance. Once this conductance becomes 
larger than the quantum conductance, the effective 
charging energy is renormalized to a value close to zero, 
and a perturbative computation in $\hat V$ breaks down. 
In this regime extended states on the cluster 
are strongly hybridized with those in the metal, 
and can be viewed as part of the extended states in the metal.

In the regime of strong electron tunneling between the substrate and cluster, 
it is not clear whether the itinerant model 
is able to produce a  Kondo effect.  
On the other hand, our local  moment cluster model gives a 
natural description  of the strong tunneling regime. In the local moment model
the localized d-electron spins can be 
viewed as a magnetic cluster embedded in the metallic 
host. This model has been analyzed in detail in 
Ref.~\onlinecite{levin}.  In our local moment model, antiferromagnetic 
exchange ($J>0$ in Eq.~(\ref{model_II_final})) between the local 
moments and the conduction electrons produces a 
Kondo effect in the strong tunneling regime, though, 
the Kondo temperature decreases very fast with increasing 
cluster size.  { We have also argued that within the local moment 
model, the effective coupling between the electrons in the 
substrate and the cluster spin must change sign as one
gradually increases the tunneling between the cluster and 
the substrate.} 

{ Both the itinerant and local moment calculations show 
that the Kondo coupling is {\em inversely} proportional to 
the {\em total spin} $S_T$ of a ferromagnetic  cluster, which 
in turn, is proportional to the size of the 
cluster.\cite{footnote} 
%This fact nicely illustrates 
%that 
The Kondo effect is due to quantum fluctuations 
%(change in $\hat z$-component) 
of the 
cluster spin, and  these are suppressed as $1/S_T$
for large ferromagnetic clusters. Thus $T_K$ goes to zero {\em exponentially}
with increasing cluster size.}

To make stronger contact with the experiments of Ref.~\onlinecite{teri} on
Co clusters on a carbon nanotube,
we also calculated the STM spectra of a
ferromagnetic cluster as described in Sec.~\ref{connection}A. 
{ We found that mesoscopic fluctuations in the charging energies
may give rise to interesting qualitative changes and asymmetries
in the STM spectrum. It is possible, for example, that 
the positively and negatively charged states of the cluster 
have very different decay rates and therefore the positive (negative) 
voltage side of the  spectrum shows  discrete levels 
while the negative (positive) voltage side 
displays a continuum spectrum.}

We can use the  model parameters extracted from the 
high energy part of the  STM spectra 
to make an estimate of $T_K$  and predict how it scales with 
cluster size.  Our results agree with the experiments 
in that they produce a rapid decrease of $T_K$ thereby rendering 
the Kondo effect impossible to observe in larger clusters.
However, the Kondo temperature we find  is already  too small by a factor of
$\sim 500$ compared to the $T_K$ observed for a small cluster of 
$\sim 8$ atoms.  
%We are
%able to obtain a $T_K$ in the experimental range provided we allow for
%experimentally resonable fluctuations in the $\delta E_{\pm,\sigma}$.
In Sec.~\ref{discussion} we have enumerated a number of effects which might
raise  $T_K$ relative to the predictions of our simple model which may be a
possible explanation for the discrepancy between theory and experiment.
We remark that by neglecting the ferromagnetic  contributions 
to the exchange coupling one would obtain a $T_K$ that is larger by a factor 
of $\sim 10^3$. Thus the ferromagnetic contributions to the effective
coupling  are essential and cannot be neglected.

Many open questions remain regarding the physics of small ferromagnetic 
clusters. Among them are: (i) Accurate estimates of the net spin of 5-50 
atom clusters supported on a substrate. 
(ii) Magnetic anistropy energies in clusters this size. (iii)
The nature of non-equilibrium and other many body effects. 
We believe that the STM 
is a crucial tool for gathering cluster-specific data for ferromagnetic 
nanoparticles and will undoubtedly reveal even more intriguing physics of
these tiny systems in the years to come.

\begin{acknowledgments}
We are very grateful to T. Odom and C.M. Lieber for many discussions on their
experiments and for sharing their data prior to publication and to   
A.H. Castro Neto and Jan von Delft, and  M. Pustilnik for useful discussions.
This research was supported by ITAMP, DIP Grant c-7.1, ISF Grant 160/01-1,  
NSF Grants No. DMR 98-09363,  DMR 99-81283 
and DMR-97-14725, and Hungarian Grants No. OTKA F030041, T029813, and 
29236. G.Z. is an E\"otv\"os  fellow. 
\end{acknowledgments}

\appendix
\section{Derivation of the Local Moment Model}
\label{sdderivation}
Consider $N_A$
magnetic impurities embedded in close proximity in a metallic host.
The Hamiltonian is $\hat H=\hat H_{\rm metal} + \hat H_{\rm int}$ where $\hat
H_{\rm metal}=\sum_{\mu, k,\sigma} \epsilon_{\mu,k}
\hat n_{\mu  k \sigma}$ 
describes the free conduction electrons and 
\begin{equation}
\hat H_{\rm int}=-\sum_{\vec r, {\vec r}'}J_F(\vec r, {\vec r}') \vec S_d(\vec r)
\cdot \vec S_d({\vec r}')+J \sum_{\vec r}\vec S_c(\vec r)
\cdot \vec S_d(\vec r)
\label{H_int}
\end{equation}
describes the direct interactions the impurity d-levels and the
interactions between the conduction electrons and the magnetic impurities.
The first term in Eq.~(\ref{H_int}) describes the ferromagnetic interaction 
among the localized d-levels of the impurity atoms and the second
term, $H_K \equiv J \sum_{\vec r}\vec S_c(\vec r) \cdot \vec S_d(\vec r)$,
describes Kondo scattering from these d-levels by the conduction
electrons.  $J_F(\vec r, {\vec r}')>0$ is the ferromagnetic exchange
interaction between two localized d-levels and $J>0$ is the bare Kondo
exchange coupling between the d-levels and the conduction
electrons.  The conduction electron spin operator at position $\vec r$ is
\begin{equation}
\vec S_c(\vec r)={1\over 2}\sum_{j,j'}\varphi_j^*(\vec r)
\varphi_{j'}(\vec r) c^\dag_{j \alpha}  
\vec \sigma_{\alpha \alpha'}c_{j' \alpha'}
\label{S_c}
\end{equation}
where $\varphi_j(\vec r)$ is the wave function of conduction
electrons with level index $j$ at $\vec r$.
The utility of Eq.~(\ref{H_int}) is two-fold: (i) It gives an expression
for the important limiting case of a ``cluster'' which consists
of just one impurity.  In this limit, $\hat H_{\rm int} =J \vec S_c(0) \cdot \vec S_d(0)$ which is just $H_K$ for a single
impurity. (ii) Eq.~(\ref{H_int}) can describe the limit
of a cluster so strongly coupled to the metallic host that the
conduction band conduction electrons of the cluster and those of the host cannot
be distinguished.  Thus, the character of the $\varphi_{j}(\vec r)$
that appear in Eq~(\ref{S_c}) vary depending on the physical situation.
For a single impurity, $\varphi_{j}(\vec r)$ is the wave function of the
host metal conduction electrons, for a cluster in the weak tunneling regime as
described in Sec.~\ref{perturbativelycoupled}  $\varphi_{j}(\vec
r)$ describes the conduction band wave functions of the ferromagnet or ferromagnetic
semiconductor and for a
cluster in the strong tunneling regime $\varphi_{j}(\vec r)$ is a
strong hybridization of the ferromagnetic conduction band electrons and the
conduction electrons of the host metal.

To derive Eq.~(\ref{model_II_final}), we
neglect anisotropy in the cluster magnetization formed by the
localized d-levels. 
We further neglect spin-wave excitations to states of different total 
d-electron spin of the cluster. In a Heisenberg-type model one may estimate 
the cost for such an excitation as
$E_{\rm spin-wave} \sim J_F (a k)^2 $ where $a$ is the lattice
spacing of the cluster's atoms and $k$ is the largest wave vector of the spin
wave allowed by the physical size of the cluster. If we take $J_F
\sim$ 0.1 eV, {\em e.g.}, then for a small cluster of size $L\sim 10$ \AA with 
lattice
constant $a\sim 2.5$ \AA,  
$k_{\rm min} \sim {\pi \over L}$ so that the minimum spin-wave energy 
$E^{\rm min}_{\rm spin-wave} \sim 60 {\rm meV}\approx 600$K.
Below that energy scale spin waves can be therefore neglected
and we can concentrate on the subspace where 
\begin{equation}
\sum_{\vec r} \vec S_d(\vec r) = N_A S_d =S_d^{\rm max}.
\end{equation}
Within the subspace of maximum total d-level spin, the d-level spin is a
rigid spin which can only change its projection on the z-axis. The
collective effect of the impurity d-levels is to give the cluster a net spin.
It is this spin that conduction electrons will scatter from--either
``directly'' in the limit of strong tunneling between the cluster and
metallic host or ``indirectly'' as described in
Sec.~\ref{perturbativelycoupled} in the limit of weak tunneling.
(In the limit of weak tunneling only conduction band electrons may hop on and
off the cluster.) 
Within the $S_d=S_d^{\rm max}$ subspace,
\begin{eqnarray}
\langle S_d,S_d'^z|\hat H_K|S_d,S_d^z\rangle & = &{J \over 2} \sum_{\vec
  r,j,j'}\varphi_j^*(\vec r) 
\varphi_{j'}(\vec r) c^\dag_{j \alpha}  
\vec \sigma_{\alpha \alpha'}c_{j' \alpha}\langle S_d,S_d'^z|\vec S_d(\vec
r)|S_d,S_d^z\rangle \\
 & \approx &  {J \over 2} \sum_{\vec
  r,j,j'}\varphi_j^*(\vec r) 
\varphi_{j'}(\vec r) c^\dag_{j \alpha}  
\vec \sigma_{\alpha \alpha'}c_{j' \alpha} {1\over N_A} 
\langle S_d,S_d'^z|\vec S_d|S_d,S_d^z\rangle.
\end{eqnarray}
The sum over $\vec r$ can be estimated,
 \begin{equation}
 \sum_{\vec r}\varphi_j^*(\vec r)
 \varphi_{j'}(\vec r)=\left\{ \begin{array}{cc}
                            1 & \mbox{$j=j'$ (normalization)}\\
                            \sim {1\over \sqrt{N_A}}  & \mbox{$j\neq j'$
 (random numbers)}
                            \end{array}
\right.
\label{off_diagonal}
\end{equation}
Neglecting the off-diagonal terms,
\begin{eqnarray}
H_K & \approx & {J\over N_A} {1\over 2} \sum_j c^\dag_{j \alpha}  
\vec \sigma_{\alpha \alpha'}c_{j' \alpha}\cdot \vec S_d \\
    & = & {J \over N_A} \vec S_c \cdot \vec S_d
\end{eqnarray}
The full Hamiltonian is
\begin{equation}
H_{\rm cluster}=\sum_{j,\sigma} \epsilon_j \hat n_{j \sigma} +{J \over
  N_A} \vec S_c \cdot \vec S_d + {E_C \over 2} (\sum_{j, \sigma}\hat
  n_{j \sigma}-n_g)^2 -\sum_{\vec r, {\vec r}'}J_F(\vec r, {\vec r}') \vec
  S_d(\vec r) \cdot \vec S_d({\vec r}')
\label{model_II}
\end{equation}
where we have put in the Coulomb charging energy by hand. 
Recall that now the $c^\dagger_{j \sigma}$ refer to conduction band
electrons, not s, p and d hybridized bands as in the itinerant model.  The
last term in~\ref{model_II} is just an irrelevant shift in the 
total energy in the
subspace $S_d=S_d^{\rm max}$ so we drop it.
Thus, we arrive at 
\begin{equation}
\hat H_{\rm cluster}=\sum_{j,\sigma} \epsilon_j \hat n_{j \sigma} +{J \over
  N_A} \vec S_c \cdot \vec S_d + {E_C \over 2} (\sum_{j, \sigma}\hat
  n_{j \sigma}-n_g)^2.
\end{equation}

\section{Matrix Elements and Kondo Couplings for the Local 
Moment Model}
\label{matrix_model_II}
In this Appendix, we include some more lengthy expressions not included in 
Sec.~\ref{model_II_coupling}.  To complete the 
evaluation the RHS of Eq.~(\ref{pert}) for the local moment model
described in Sec.~\ref{model_II_coupling} we need the matrix elements:
\begin{eqnarray}
M_2&\equiv& 
{}^N\langle S_T,S_T-1|c_{j \uparrow}|S_T+1/2,S_T-1/2\rangle_j^{N+1}
 \nonumber \\
&= & - \sum_{S_c^z=-S_c}^{S_c} \sqrt{{S_c -S_c^z\over 2 S_c}}
\langle S_c,S_d;S_c^z,S_T-S_c^z-1|S_T,S_T-1\rangle \nonumber \\
&& \langle S_c-1/2,S_d;S_c^z+1/2,S_T-S_c^z-1|S_T+1/2,S_T-1/2\rangle
\label{m_2}
\end{eqnarray}
\begin{eqnarray}
M_3&\equiv& 
{}_j^{N-1}\langle S_T+1/2,S_T-1/2|c_{j \uparrow}|S_T,S_T\rangle^N
\nonumber \\
& = &\sum_{S_c^z=-S_c}^{S_c} \sqrt{{S_c +S_c^z\over 2 S_c}}
\langle S_c,S_d;S_c^z,S_T-S_c^z|S_T,S_T\rangle \nonumber \\
&&\langle S_c-1/2,S_d;S_c^z-1/2,S_T-S_c^z|S_T+1/2,S_T-1/2\rangle
\label{m_3}
\end{eqnarray}
\begin{eqnarray}
M_4&\equiv& 
{}^N\langle S_T,S_T-1|c_{j \downarrow}^\dagger|S_T+1/2,S_T-1/2\rangle_j^{N-1}
\nonumber \\
&=& \sum_{S_c^z=-S_c}^{S_c} \sqrt{{S_c -S_c^z\over 2 S_c}}
\langle S_c,S_d;S_c^z,S_T-S_c^z-1|S_T,S_T-1\rangle \nonumber \\
&&\langle S_c-1/2,S_d;S_c^z+1/2,S_T-S_c^z-1|S_T+1/2,S_T-1/2\rangle.
\label{m_4}
\end{eqnarray}

The matrix elements then directly yield the expression for the generalized
Kondo couplings for the local moment model:
\begin{equation}
\tilde J^{\mu \nu}_{s} = \sqrt{{2 \over S_T}} \sum_{j=I+1}^A {V_\mu^{j,k_f}}^* V_\nu^{j,k_i}\Biggl(
{M_4 M_3 \over \delta E_{-,\uparrow}+\epsilon_A -\epsilon_j}
-
{M_2 M_1 \over  \delta E_{+,\uparrow} -\epsilon_{I+1} +\epsilon_j}  
  \Biggr), 
\label{J_II}
\end{equation}
where $M_1$ denotes the matrix element already given in Eq.~(\ref{m_1}).
The products $M_4 M_3$ and $M_2 M_1$ can be evaluated directly:
\begin{equation}
\tilde J^{\mu \nu}_{s} = -{1 \over S_T + 1} \sum_{j=I+1}^A {V_\mu^{j,k_f}}^* V_\nu^{j,k_i}\Biggl(
{1 \over \delta E_{-,\uparrow}+\epsilon_A -\epsilon_j}
+
{1 \over  \delta E_{+,\uparrow} -\epsilon_{I+1} +\epsilon_j}  
  \Biggr), 
\label{eq:J_II_final}
\end{equation}
giving a ferromagnetic contribution when $J>0$.  If $J<0$ and $S_T=S_d + S_c$ 
($\uparrow \to \downarrow$)
one finds $-{1 \over S_T +1} \to {1\over S_T}$ which agrees with Eq.~(\ref{J_s}) of the itinerant model with $S_T \to S_0$.

A calculation similar to the one that led to Eq.~(\ref{J_II}) and 
Eq.~(\ref{eq:J_II_final}) then yields:
\begin{equation}
\tilde J^{\mu \nu}_{e} = \sum_{j>A} {V_\mu^{j,k_f}}^* V_\nu^{j,k_i}\Biggl(
{ {2 \over 2 S_T +1} \over \delta E_{+,\downarrow}-\epsilon_{A+1} + \epsilon_j}
-
{ {1\over S_T +1} \Bigl( {2 S_c \over 2S_c +1}\Bigr)  \over \delta E_{+,\uparrow}+\Delta_s -\epsilon_{A+1} +\epsilon_j}  
  \Biggr), 
\end{equation}
and from the doubly occupied states
\begin{equation}
\tilde J^{\mu \nu}_{d} = \sum_{j<=I} {V_\mu^{j,k_f}}^* V_\nu^{j,k_i}\Biggl(
{{2 \over 2 S_T +1}  \over \delta E_{-,\downarrow} +\epsilon_I - \epsilon_j}
-
{ {1\over S_T +1} \Bigl( {2 S_c \over 2S_c +1}\Bigr) \over \delta E_{-,\uparrow}+\Delta_s +\epsilon_I -\epsilon_j}  
  \Biggr) 
\end{equation}
where we have agained assumed $J>0$, giving
$\tilde J^{\mu \nu}_{e}>0$ and $\tilde J^{\mu \nu}_{d}>0$ since 
$\delta E_{\pm,\sigma}\sim E_C/2$. 
In the limit of a 
single point of contact, one recovers an expression similar 
to Eq.~(\ref{eq:Jestimate}), Eq.~(\ref{eq:Jtilde_rho}), except with an 
overall sign difference when $J>0$
(and the precise form of the mesoscopic fluctuations).
In the case $J<0$ (take $\uparrow \to \downarrow$ and $\downarrow \to 
\uparrow$) we have 
${2 \over 2 S_T +1} \to -{2 \over 2 S_T+1}$ and 
${1\over S_T +1} \Bigl( {2 S_c \over 2 S_c +1}\Bigr) \to -{1\over S_T} \Bigl( {2 S_c \over 2S_c +1}\Bigr)$
which makes $\tilde J^{\mu \nu}_{e}<0$ and $\tilde J^{\mu \nu}_{d}<0$.

The reduced matrix elements\cite{sakurai} for the all the states of
the local moment model are given below:
\subsection{J$>$0}
\begin{eqnarray}
^{N-1}_j\langle S_T+1/2||c_{j}||S_T\rangle_d^N
 \nonumber 
& = & \sqrt{ 2 S_c \over 2 S_c +1} \sqrt{2S_T +2}
\end{eqnarray}
\begin{eqnarray}
^{N-1}_j\langle S_T-1/2||c_{j}||S_T\rangle_d^N
 \nonumber 
& = & {2 S_T \over \sqrt{2S_T +1}}
\end{eqnarray}
\begin{eqnarray}
^{N+1}_j\langle S_T+1/2||c_{j}^\dagger||S_T\rangle_s^N
 \nonumber 
& = & {2 S_T +1  \over \sqrt{2S_T +2}}
\end{eqnarray}
\begin{eqnarray}
^{N-1}_j\langle S_T+1/2||c_{j}||S_T\rangle_s^N
 \nonumber 
& = & \sqrt{2S_T +2}
\end{eqnarray}
\begin{eqnarray}
^{N+1}_j\langle S_T+1/2||c_{j}^\dagger||S_T\rangle_e^N
 \nonumber 
& = & {2 S_T +1  \over \sqrt{2S_T +2}} \sqrt{{2S_c \over 2S_c +1}}
\end{eqnarray}
\begin{eqnarray}
^{N+1}_j\langle S_T-1/2||c_{j}^\dagger||S_T\rangle_e^N
 \nonumber 
& = & \sqrt{2S_T +1}
\end{eqnarray}

\subsection{J$<$0}

\begin{eqnarray}
^{N-1}_j\langle S_T+1/2||c_{j}||S_T\rangle_d^N
 \nonumber 
& = & { 2 S_T+2 \over \sqrt{2S_T +1}}
\end{eqnarray}
\begin{eqnarray}
^{N-1}_j\langle S_T-1/2||c_{j}||S_T\rangle_d^N
 \nonumber 
& = & \sqrt{2 S_T}\sqrt{2S_c \over 2S_c +1}
\end{eqnarray}
\begin{eqnarray}
^{N+1}_j\langle S_T-1/2||c_{j}^\dagger||S_T\rangle_s^N
 \nonumber 
& = & {2 S_T +1  \over \sqrt{2S_T}}
\end{eqnarray}
\begin{eqnarray}
^{N-1}_j\langle S_T-1/2||c_{j}||S_T\rangle_s^N
 \nonumber 
& = & \sqrt{2S_T}
\end{eqnarray}
\begin{eqnarray}
^{N+1}_j\langle S_T+1/2||c_{j}^\dagger||S_T\rangle_e^N
 \nonumber 
& = & \sqrt{2 S_T +1} 
\end{eqnarray}
\begin{eqnarray}
^{N+1}_j\langle S_T-1/2||c_{j}^\dagger||S_T\rangle_e^N
 \nonumber 
& = & { 2S_T +1 \over \sqrt{2S_T}}\sqrt{{2S_c \over 2S_c+1}}
\end{eqnarray}

%\begin{figure}
%\centerline{\epsfig{figure=stm_mod.eps, width= 4 in}}
%\caption{
%Experimental setup to realize a tunable single tunneling mode situation. 
%The insulating oxide layer should be thick enough to 
%prevent the formation of the Kondo effect with electrons in 
%the support, but it should be thin enough  
%to give a measurable conductance.  The  Kondo effect is formed 
%between  electrons on the STM tip  and the magnetic moment of the 
%ferromagnetic cluster. 
%\label{stm}}
%\end{figure}

%\begin{figure}
%\centerline{\epsfig{figure=stm_subnanometer_2.eps, width= 4 in}}
%\caption{Computed STM spectrum of a sub-nanometer ferromagnetic cluster 
%with parameters $N_A=7$, $E_C^+=.05{\rm eV}$, $E_C^-=.25{\rm eV}$, 
%$\Gamma_0=.08{\rm eV}$, $d_0 = -0.12 {\rm eV}$,
% and  $\mu=0$ at  4 K.  Two spectra are plotted that overlay one another except
%near zero bias.  The lighter line-dot curve is the spectrum with Kondo peak 
%restored following  Ref.~\cite{ujsaghy} 
%using the experimental  Kondo temperature $T_K\sim 80$ K.\cite{teri}  
%The darker plain curve is our calculation with $T_K<<$ 4 K, thus no Kondo
%peak is present.} 
%\label{subnanometer_spectrum}
%\end{figure}


\begin{thebibliography}{99}
\bibitem{quantumdot} In a semiconductor quantum dots\cite{david,leo}
  and nanotube quantum dots\cite{cobden} $n_g$ can be tuned by a variable
  external gate voltage, thereby controlling the equilibrium number of 
electrons on the dot (cluster).  For a ferromagnetic cluster adsorbed on a
  metallic surface, $n_g$ is set by the chemistry of the cluster-metal
  interface and can not be freely changed in an experiment.
\bibitem{anisotropy_comment} The itinerant model presented has a
  ($2S_0+1$)-fold degeneracy.  In a real
  ferromagnetic cluster there will be anisotropy which, in the
  simplest case, will appear as a uniaxial anistropy\cite{kleff}. 
  For Co, the anistropy energy can be estimated to be of the order\cite{kleff}
 .01-.1 meV $\approx$ .1-1.0 K which is safely below the experimental
  temperature range of 4-100 K\cite{teri} so we ignore it.
\bibitem{Glazman} 
An { alternative}  proof  was given independently 
for the special case of a $2 \times 2$ $J^{\mu\nu}$  
in Ref.~\cite{Glazmanpaper}.
\bibitem{remark} 
Using the experimentally observed level widths of the 
Co clusters  on nanotubes\cite{teri} and the density of states
of a metallic nanotube\cite{dresselhaus} we estimate the bare hopping
element (overlap integral) in a tight binding model from an atom of the 
nanotube to the nearest Co cluster atom to be $\sim$3-4 eV. This 
is comparable to the overlap integral of the nearest neighbor C-C overlap
integral in nanotubes: 2.5 eV.\cite{dresselhaus}
\bibitem{footnote}
{ We assumed  that the magnetization per 
atom on the cluster is independent of the size of the cluster.}
\end{thebibliography}
\end{document}